\documentclass[12pt,a4paper]{article}
\pdfoutput=1
\usepackage{graphicx}
\usepackage[T1]{fontenc}
\usepackage[sc,osf]{mathpazo}
\usepackage{a4wide}  
\usepackage{slashed}
\usepackage{latexsym,amsthm,amsfonts,amsmath,mathrsfs,amssymb}
\usepackage{booktabs} 
\usepackage[unicode,implicit]{hyperref}
\hypersetup{%
  pdftitle    = {The structure of all the supersymmetric solutions of ungauged N=(1,0),d=6 supergravity}
  pdfkeywords = {supergravity, supersymmetry, solutions, BPS, ungauged, unbroken},
  pdfauthor   = {Pablo A. Cano and Tomas Ortin},
  plainpages  = true,
  colorlinks  = true,
  citecolor   = blue,
  urlcolor    = red,
  linkcolor   = black
}
\newcommand{\hepth}[1]{{\tt
\href{http://www.arXiv.org/abs/hep-th/#1}{hep-th/#1}}}

\newcommand{\arxiv}[1]{{\tt arXiv:\href{http://www.arXiv.org/abs/#1}{#1}}}
\newcommand{\doi}[1]{{\tt DOI:\href{http://dx.doi.org/#1}{#1}}}

\makeatletter
\@addtoreset{equation}{section}
\makeatother

\begin{document}

\begin{flushright}
\small
IFT-UAM/CSIC-18-040\\
\texttt{arXiv:1804.04945 [hep-th]}\\
April 13\textsuperscript{th}, 2018\\
\normalsize
\end{flushright}

\vspace{0cm}

\begin{center}

  {\Large {\bf {The structure of all the supersymmetric solutions\\[.5cm]
        of ungauged $\mathcal{N} = (1,0),d=6$ supergravity}}}
 
\vspace{2.5cm}

\renewcommand{\thefootnote}{\alph{footnote}}
{\sl\large Pablo A.~Cano$^{1}$}${}^{,}$\footnote{E-mail: {\tt pablo.cano [at] uam.es}}
{\sl\large and Tom\'{a}s Ort\'{\i}n$^{1}$}${}^{,}$\footnote{E-mail: {\tt Tomas.Ortin [at] csic.es}},

\setcounter{footnote}{0}
\renewcommand{\thefootnote}{\arabic{footnote}}

\vspace{1.5cm}

${}^{1}${\it Instituto de F\'{\i}sica Te\'orica UAM/CSIC\\
C/ Nicol\'as Cabrera, 13--15,  C.U.~Cantoblanco, E-28049 Madrid, Spain}\\ 

\vspace{2.5cm}


{\bf Abstract}

\end{center}

\begin{quotation}
  We characterize all the supersymmetric configurations and solutions of
  minimal ($\mathcal{N}=(1,0)$) $d=6$ supergravity coupled in the most general
  gauge-invariant way to an arbitrary number of tensor and vector multiplets
  and hypermultiplets.
\end{quotation}

\newpage
\pagestyle{plain}

\tableofcontents

\section*{Introduction}

After the pioneering works of Gibbons, Hull and Tod
\cite{Gibbons:1982fy,Tod:1983pm} on the characterization of the supersymmetric
solutions of pure (minimal) $\mathcal{N}=2,d=4$ supergravity, a great effort
(leading to a wealth of very important and useful results we will not try to
review here) has been devoted to the characterization
(\textit{a.k.a.}~''classification'') of the supersymmetric configurations and
solutions of more general supergravity theories. This effort has been
particularly intense and fruitful in the realm of the so-called
$\mathcal{N}=2$ supergravity theories in $d=4,5$ and $6$ dimensions. These are
theories with 8 supercharges that admit timelike supersymmetric solutions,
which include black holes, and not just null supersymmetric solutions, which
in 4 dimensions only include waves and ``stringy cosmic strings''. These
theories admit many different matter couplings but the amount of supersymmetry
they have constrains their structure the right amount which is needed to endow
them with interesting geometries and dualities.

In $d=4$, using the ``bilinear method'' of Ref.~\cite{Gauntlett:2002nw}, the
timelike case of the most general $\mathcal{N}=2$ theory was worked out in
Ref.~\cite{Meessen:2012sr}, culminating a long series of works in which
theories with more general matter couplings were studied
\cite{Caldarelli:2003pb,Meessen:2006tu,Huebscher:2006mr,Cacciatori:2008ek,Hubscher:2008yz,Klemm:2009uw,Klemm:2010mc},
in which the the null case (in absence of non-Abelian gaugings) was also
solved. Only the null case for the most general non-Abelian-gauged theory
remains to be worked out.

In $d=5$ dimensions, the timelike and null cases have been solved for the most
general theory after another long sequence of works dealing with increasingly
complicated matter couplings
\cite{Gauntlett:2002nw,Gauntlett:2003fk,Gutowski:2004yv,Gauntlett:2004qy,Gutowski:2005id,Bellorin:2006yr,Bellorin:2007yp,Bellorin:2008we}.
An interesting aspect of the 5-dimensional case is that the supersymmetric
solutions that admit an isometry which acts with no fixed points can be
reduced to a supersymmetric solution of a $\mathcal{N}=2,d=4$
supergravity. 


Less is known about the six-dimensional case, which only admits null
Killing spinors, which nonetheless has interesting applications.  For
example, supersymmetric solutions of $d=6$ supergravity have recently
proven to be very useful in the context of the fuzzball proposal
\cite{Mathur:2005zp}, thanks to the construction horizonless {\it
  microstate geometries}
\cite{Bena:2011dd,Giusto:2013rxa,Bena:2015bea,deLange:2015gca} that
are able to account for a finite fraction of the black hole entropy.

In spite of the interest of this case, in $d=6$ dimensions, however, only two
theories have been completely studied so far: pure supergravity, in
Ref.~\cite{Gutowski:2003rg}, and Fayet-Iliopoulos-gauged supergravity coupled
to some vector multiplets in Ref.~\cite{Cariglia:2004kk}. In between these two
theories, there is a huge gap corresponding to ungauged theories coupled to
arbitrary numbers of vector and tensor multiplets and also to hypermultiplets
which are the theories we are going to consider here. This should be
understood as a first step towards the complete characterization of all the
supersymmetric solutions of the most general matter-coupled $N=(1,0),d=6$
supergravities: on the one hand, only if one considers vector multiplets can
one gauge any symmetries of the ungauged theory. On the other hand, the
symmetries that can be gauged are $R$-symmetry, which is gauged via
Fayet-Iliopoulos terms (the case considered in Ref.~\cite{Cariglia:2004kk})
and the isometries of the hyperscalar manifold and of the scalar manifold
associated to the tensor multiplets. Thus, if one wants to take a step beyond
what is already known, one is forced to consider, at least, tensor and vector
multiplets together.

There is another reason for considering these two kinds of multiplets
simultaneously: the existence of duality between ungauged theories with tensor
and vector multiplets compactified in a circle discovered in
Ref.~\cite{Cano:2016rls}. More precisely, in that reference it was shown that
the dimensional reduction of one of these theories with just one tensor
multiplet ($n_{T}=1$) and an arbitrary number of vector multiplets $n_{V}$ and
that of a theory with $n_{T}'=1+n_{V}$ and $n_{V}'=0$ give exactly the same
$\mathcal{N}=1,d=5$ theory with $n_{V5}=n_{V}+2$ vector multiplets. 

This situation is entirely analogous to the identity between the reductions on
dual circles of the $\mathcal{N}=2A,d=10$ and $\mathcal{N}=2B,d=10$ theories
\cite{Bergshoeff:1995as} which signals, at the effective action level, the
T-duality between type~IIA and IIB superstring theories discovered in
Refs.~\cite{Dai:1989ua,Dine:1989vu}. In the case at hands, there is no known
stringy/brany duality underlying the duality found at the supergravity level.
Nevertheless, it is possible to derive a set of Buscher-type duality rules
that transform solutions of the $n_{T}=1,n_{V}$ theory with an isometry into a
solutions of the $n_{T}'=n_{V}+1,n_{V}'=0$ theory and vice-versa. 

Although this duality has been found in the bosonic equations of motion it is,
most likely, a duality between the two complete supergravity theories and,
therefore, it is to be expected that the supersymmetric solutions of both
kinds of theories are related by it. A first step to check whether this is
true is the characterization of all the solutions of the ungauged theories
with arbitrary numbers of vector and tensor multiplets, which we are going to
present here. The relation between the supersymmetric solutions will be
studied elsewhere.

Early work on supersymmetric solutions of the theories that we are going to
consider here can be found in Ref.~\cite{Duff:1996cf}, but this work is quite
far from the systematic and exhaustive approach we aim to pursue here. More
recently, M.~Akyol and G.~Papadopoulos in Ref.~\cite{Akyol:2010iz} solved the
Killing Spinor Equations of these theories in the most general, gauged case
identifying the geometry and the field strengths of supersymmetric field
configurations. However, since the main goal of that paper was to study the
different kinds of Killing spinors admissible by the supersymmetric
configurations, they did not solve the Bianchi identities of the vector field
strength nor did they impose the equations of motion on them. Thus, the
supersymmetric configurations were not completely characterized and the
supersymmetric solutions (the equations that they have to solve) were left
unidentified.

It is known that supersymmetry ensures that some of the equations of motion of
supersymmetric solutions are related among them or automatically solved but
there is always a number of them which are independent and need to be
solved. The relations between the equations of motion of supersymmetric field
configurations can be obtained bia the so-called Killing Spinor Identities
(KSI) \cite{Kallosh:1993wx,Bellorin:2005hy} or via the integrability
conditions of the Killing Spinor Equations. In order to construct the KSIs one
needs the locally supersymmetric action of the theory, which does not exist
for the theories under consideration because they include 2-forms with
(anti-)self-dual 3-form field strengths.\footnote{It might be possible to
  derive the KSIs from the pseudo-action Eq.~(\ref{eq:pseudoaction}), but this
  requires further investigation.} Thus, in this work we will find these
relations from the integrability conditions of the KSEs.  We will identify
the independent and non-trivial (for supersymmetric configurations) equations
of motion and we will impose them, together with the Bianchi identities of the
vector fields, on the supersymmetric configurations, finding a reduced number
of simplified differential equations to be solved. Our main result, summarized
in Section~\ref{sec-summary}, will be this set of simplified differential
equations and a recipe to construct supersymmetric solutions of theories with
an arbitrary number of vector and tensor multiplets.\footnote{After this work
  was already completed we learned about another work by H.~het Lam and
  S.~Vandoren \cite{Lam:2018jln} which studies the case of coupling to an
  arbitrary number of tensor multiplets only from the same point of view.}  We
will also study some further simplifications of these equations for particular
cases and make contact with the results on classifications of
$\mathcal{N}=1,d=5$ supersymmetric solutions, partially confirming the results
of this paper.\footnote{The detailed comparison of our results with those of
  Ref.~\cite{Akyol:2010iz} is very complicated because we are not solving
  exactly the same problem: we consider solutions with \textit{at least} one
  unbroken supersymmetry and, in Ref.~\cite{Akyol:2010iz}, field
  configurations with \textit{exactly} one, two etc.~unbroken supersymmetries
  are considered. Furthermore, we make some explicit choices of coordinates
  that lead to the definition of the, most useful, \textit{base space}, in
  which many of the objects that we determine are defined. The definition of
  base space is not used in Ref.~\cite{Akyol:2010iz}. One can only compare the
  components of the 3-form field strengths and some general structures of the
  supersymmetric solutions and check that, indeed, they agree.}

This paper is organized as follows: in Section~\ref{sec-sugra} we present the
theories, their field content, (pseudo-)action, equations of motion and
supersymmetry transformation rules. In Section~\ref{sec-configurations} we
characterize the field configurations which, satisfying the equations of
motion or not, admit at least one Killing spinor.  In
Section~\ref{sec-solutions} we impose the equations of motion on the
supersymmetric configurations we have characterized. Only a few of them are
actually independent and, therefore, as usual, the number of equations that
have to be solved by the building blocks of a supersymmetric configuration is
very reduced. At this stage, we have achieved all the goals we were aiming for
and is simply rest to summarize our results. We do this, and conclude, in
Section~\ref{sec-summary}.

\section{Ungauged six-dimensional supergravity}
\label{sec-sugra}

Six-dimensional supergravity coupled to matter has been described in
increasing levels of generality in
Refs.~\cite{Nishino:1986dc,Sagnotti:1992qw,Nishino:1997ff,Ferrara:1997gh,Riccioni:1997np,Riccioni:1999xq,Riccioni:2001bg}.
For the sake of completeness we review here the most relevant results.  We
will mostly use the notation of \cite{Riccioni:2001bg}.  The bosonic sector of
six-dimensional supergravity contains the graviton, represented by the
Vielbein $e^{a}{}_{\mu}$, a number $n_{T}$ of scalars
$\varphi^{\underline\alpha}$, $n_{T}+1$ two-forms, $B^{r}_{\mu\nu}$, with
respective field strengths

\begin{equation}
\label{eq:Hrdef}
H^{r}=dB^{r}+\tfrac{1}{2}c^{r}_{\ ij}F^{i}\wedge A^{j},
\end{equation}

\noindent
 $n_{V}$ abelian
vectors $A^{i}$, with field strengths 

\begin{equation}
F^{i}=dA^{i}, 
\end{equation}

\noindent
and $4n_{H}$ hyperscalars $\phi^{X}$. The fermionic sector consists of the
gravitino $\psi^{A}_{\mu}$, $n_{T}$ tensorinos $\chi^{M\ A}$, $n_{V}$ gauginos
$\lambda^{i\ A}$ and $2 n_{H}$ hyperinos $\Psi^{\bf{a}}$. Let us explain how
these fields couple among them.


The scalars $\varphi^{\underline{\alpha}}$, $\underline{\alpha}=1,...,n_{T}$,
parametrize the coset $\mathrm{SO}(n_{T},1)/\mathrm{SO}(n_{T})$. The indices
$M, N=1,...,n_{T}$ belong to the fundamental representation of
$\mathrm{SO}(n_{T})$, while $r,s=0,1,...,n_{T}$ label the fundamental
representation of $\mathrm{SO}(n_{T},1)$. On the other hand, $A=1,2$ is an
$\mathrm{Sp}(1)$ index. Let us introduce a coset representative of
$\mathrm{SO}(n_{T},1)/\mathrm{SO}(n_{T})$ as a $(n_{T}+1)\times (n_{T}+1)$
matrix, $L_{r}{}^{s}$, which belongs to $\mathrm{SO}(n_{T},1)$. It is useful
to split its components in the form $L_{r}\equiv L_{r}{}^{0}$ and
$L_{r}{}^{M}$, so that they satisfy

\begin{equation}\label{Eq:eta}
L_{r}L_{s}-L_{r}{}^{M}L_{s}{}^{M}=\eta_{rs},
\end{equation}

\noindent
where $\eta_{rs}=\operatorname{diag}(+1,-1,-1,...,-1)$. The indices $r, s$ are
lowered and raised with $\eta_{rs}$ ($\eta^{rs}$) in the way
$L^{r}=\eta^{rs}L_{s}$, $L^{rM}=-\eta^{rs} L_{r}{}^{M}$. We also have the
relations

\begin{equation}
L^{r}L_{r}=1, 
\quad 
L^{r}L_{r}{}^{M}=0, 
\quad L_{r}{}^{M}L^{r N}
=
\delta^{MN}.
\end{equation}

We will also need the following symmetric but not constant tensor:

\begin{equation}\label{Eq:Grs}
G_{rs}=L_{r}L_{s}+L_{r}{}^{M}L_{s}{}^{M}.
\end{equation}

The scalars parametrize the coset representative, and we have the following
relations

\begin{eqnarray}
\partial_{\underline{\alpha}}L_{r}
&=&
V_{\underline{\alpha}}{}^{M}L_{r}{}^{M},
\\
& & \nonumber \\
\partial_{\underline{\alpha}}L_{r}{}^{M}
&=&
-A_{\underline{\alpha}}{}^{M}{}_{N}L_{r}{}^{N}+V_{\underline{\alpha}}{}^{M}L_{r},
\end{eqnarray}

\noindent
where $\partial_{\underline{\alpha}}\equiv\partial/\partial
\varphi^{\underline{\alpha}}$ and $V_{\underline{\alpha}}{}^{M}$ is the
Vielbein, which satisfies

\begin{equation}
V_{\underline{\alpha}}{}^{M}V_{\underline{\beta}}{}^{M} 
=
g_{\underline{\alpha}\underline{\beta}},
\end{equation}

\noindent
where $g_{\underline{\alpha}\underline{\beta}}$ is the metric of the scalar
manifold associated to the tensor multiplets.

It is also convenient to define the 3-forms field strengths in the basis
defined by the coset $\mathcal{H},\mathcal{H}^{M}$, because the supersymmetric
transformation rules are written are written in terms of them. They are
related to those defined in Eq.~(\ref{eq:Hrdef}) by

\begin{equation}
\mathcal{H}=L_{r}H^{r}, 
\quad 
\mathcal{H}^{M}=L_{r}{}^{M}H^{r}.
\end{equation}

In this way, we can distinguish the several supermultiplets of this theory: we
have the supergravity multiplet $\{e^{a}{}_{\mu}, \psi^{A}_{\mu},
\mathcal{H}\}$, $n_{T}$ tensor supermultiplets $\{\chi^{M\ A},
\varphi^{\underline{\alpha}}, \mathcal{H}^{M}\}$, $n_{V}$ vector supermultiplets
$\{A^{i}, \lambda^{i\ A}\}$ and $n_{H}$ hypermultiplets $\{ \phi^{X},
\Psi^{\bf{a}}\}$.

On the other hand, the hyperscalars $\phi^{X}$ parametrize a
quaternionic-Kahler manifold of holonomy $\mathrm{Sp}(1)\times
\mathrm{Sp}(n_{H})$. For completeness, we give here certain formulas which
are used throughout the calculations in the text. The Vielbein of the
quaternionic manifold is denoted as $V^{{\bf{a}}A}{}_{X}$, with $X=1,\ldots
4n_{H}$, while ${\bf{a}}$ and $A$ are indices of $\mathrm{Sp}(n_{H})$ and
$\mathrm{Sp}(1)$, respectively. We are interested in the $\mathrm{Sp}(1)$
connection, which is denoted $A_{X}{}^{A}{}_{B}$ and it is anti-hermitian in
the $A,B$ indices. Equivalently, we can write the components of the connection
in the adjoint representation

\begin{equation}
A_{X}{}^{A}{}_{B}
=
\tfrac{i}{2}(\sigma^{x})^{A}{}_{B}A^{x}{}_{X}
\,\,\,\,\Leftrightarrow \,\,\,\,
A^{x}{}_{X}
=-i(\sigma^{x})^{A}{}_{B}A_{X}{}^{B}{}_{A},
\end{equation}

\noindent
where $x=1,2,3$ and $\sigma^{x}$ are the Pauli matrices.
The field-strength of the connection is defined as

\begin{equation}
\mathcal{F}_{XY}{}^{A}{}_{B}
=
\partial_{X}A_{Y}{}^{A}{}_{B}-\partial_{Y}A_{X}{}^{A}{}_{B}
+\left[A_{X},A_{Y}\right]^{A}{}_{B}.
\end{equation}

We also have the following relations
\cite{Bagger:1983tt,Riccioni:2001bg}\footnote{We thank S.J.G.~Vandoren for
  pointing to us a missprint in the first of these relations in
  Ref.~\cite{Bagger:1983tt}.}

\begin{equation}
\begin{aligned}
V_{{\bf{a}}A}{}^{X}V^{{\bf{b}}AY}
+V_{{\bf{a}}A}{}^{Y}V^{{\bf{b}}AX}
&=
g^{XY}\delta^{\bf{b}}_{\bf{a}},
\\
& \\
V_{{\bf{a}}A}{}^{X}V^{{\bf{a}}BY}
+V_{{\bf{a}}A}{}^{Y}V^{{\bf{a}}BX} 
&=
g^{XY}\delta^{B}_{A},\\
& \\
V_{{\bf{a}}AX}V^{{\bf{a}}}{}_{BY}+V_{{\bf{a}}BX}V^{{\bf{a}}}{}_{AY}
&=
\mathcal{F}_{XYAB}.
\end{aligned}
\label{VVrel}
\end{equation}

The quaternionic structures are related to the field strength according to

\begin{equation}
\mathcal{J}^{xX}{}_{Y}
=
\tfrac{1}{2}\mathcal{F}^{xX}{}_{Y}
\equiv
-\tfrac{i}{2}(\sigma^{x})^{B}{}_{A}\mathcal{F}^{X}{}_{Y}{}^{A}{}_{B}.
\label{Jdef}
\end{equation}

\noindent
They are covariantly constant with respect to the $\mathrm{Sp}(1)$ connection
and they satisfy the quaternionic algebra

\begin{equation}
\mathcal{J}^{x}\cdot\mathcal{J}^{y}
=
-\delta^{xy}+\epsilon^{xyz}\mathcal{J}^{z}.
\end{equation}

\subsection{Field equations and supersymmetry transformations}

It is always most convenient to have an action principle from which the
equations of motion can be derived.  Precisely, one of the difficulties of
$\mathcal{N}=(1,0),d=6$ supergravity is that it contains self-dual 3-forms
whose equations of motion cannot be obtained from a covariant action
functional unless one introduces PST-type auxiliary variables
\cite{Pasti:1995tn,Pasti:1996vs,Pasti:1997gx,Bandos:1997ui} and reformulates
the theory using them.  However, one can also use a ``pseudo-action'' (which
is not supersymmetric) from which, through its functional derivatives, 
one obtains equations that have to be supplemented by the duality constraints.
For instance, this has been done in
Refs.~\cite{Bergshoeff:1995sq,Meessen:1998qm} for the $\mathcal{N}=2B,d=10$
theory, whose Ramond-Ramond 4-form has a self-dual 5-form field strength, and
for the case at hands (with no hypermultiplets) in
Ref.~\cite{Cano:2016rls}. The pseudo-action we need is\footnote{We follow the
  conventions of \cite{Ortin:2015hya}.}  \cite{Riccioni:2001bg}

\begin{equation}
\label{eq:pseudoaction}
\begin{aligned}
S=\frac{1}{16\pi G_{N}^{(6)}}\int d^{6}x\sqrt{|g|}
\bigg\{
&
R-\partial_{\mu}L^{r}\partial^{\mu}L_{r}
+\tfrac{1}{3}G_{rs}H^{r}{}_{\mu\nu\rho}H^{s\ \mu\nu\rho}
-L_{r}c^{r}{}_{ij}F^{i}_{\mu\nu}F^{j\ \mu\nu}\\
& \\
&
-\tfrac{1}{4}c_{r\ ij}
\frac{\epsilon^{\mu\nu\rho\sigma\lambda\eta}}{\sqrt{|g|}}
B^{r}{}_{\mu\nu}F^{i}_{\rho\sigma}F^{j}_{\lambda\eta}
+2g_{XY}\partial_{\mu}\phi^{X}\partial^{\mu}\phi^{Y}
\bigg\},
\end{aligned}
\end{equation}

\noindent
and has to be supplemented by the self-duality relations 

\begin{equation}
\star G_{rs} H^{r}=-\eta_{rs}H^s.
\end{equation}

Given the relations Eqs.~(\ref{Eq:Grs}) and (\ref{Eq:eta}), these equations
imply that the rotated field strengths $\mathcal{H}$ and $\mathcal{H}^{M}$
are, respectively, anti-self-dual and self-dual

\begin{equation}
\mathcal{H}^{+}=0,\quad \mathcal{H}^{M-}=0.
\end{equation}

The theory is invariant under the gauge transformations

\begin{equation}
A^{i}\rightarrow A^{i}+d\Lambda^{i}, 
\quad 
B^{r}\rightarrow B^{r}
-\tfrac{1}{2}c^{r}{}_{ij}A^{i}\wedge d\Lambda^{j}+d\chi^{r},
\end{equation}

\noindent
for arbitrary $0-$ and $1-$forms $\Lambda^{i}$ and $\chi^{r}$, providing that
the constants $c^{r}{}_{ij}$ satisfy the relation

\begin{equation}
\eta^{rs}c_{r\ (ij|}c_{s\ |k)l}=0\, ,
\end{equation}

\noindent
which we will assume to hold. Then the field equations are gauge-invariant and
they read

\begin{eqnarray}
\label{Eq:EOM}
\mathcal{E}_{\mu\nu}
& = & 
R_{\mu\nu}+g_{\underline{\alpha}\underline{\beta}}\partial_{\mu}\varphi^{\underline{\alpha}}\partial_{\nu}\varphi^{\underline{\beta}}
+G_{rs}H^{r}{}_{\mu\rho\sigma}H^{s}{}_{\nu}{}^{\rho\sigma}
-2L_{r}c^{r}{}_{ij}F^{i}{}_{\mu\rho}F^{j}{}_{\nu}{}^{\rho}
\nonumber \\
& & \nonumber \\
& &
+\tfrac{1}{4}g_{\mu\nu}L_{r}c^{r}{}_{ij}
F^{i}{}_{\rho\sigma}F^{j\  \rho\sigma}
+2g_{XY}\partial_{\mu}\phi^{X}\partial_{\nu}\phi^{Y},
\\
& & \nonumber \\
\mathcal{E}_{r} 
& = & 
d\left(\star G_{rs}H^{s}\right)+\tfrac{1}{2}c_{r\ ij}F^{i}\wedge F^{j},
\\
& & \nonumber \\
\mathcal{E}_i 
& = & 
d(\star L^{r}c_{r\ ij}F^{j})-2c_{r\ ij}H^{r}\wedge F^{j},
\\
& & \nonumber \\
\mathcal{E}^{\underline{\alpha}} 
& = & 
\mathfrak{D}_{\mu}\partial^{\mu}\varphi^{\underline{\alpha}}
-\tfrac{2}{3}V^{\underline{\alpha} M}\mathcal{H}^{M}{}_{\mu\nu\rho}
\mathcal{H}^{\mu\nu\rho}
+\tfrac{1}{2}V^{\underline{\alpha} M}L_{r}{}^{M}c^{r}{}_{ij}
F^{i}{}_{\mu\nu}F^{j\ \mu\nu},
\\
& & \nonumber \\
\mathcal{E}^{X} 
& = & 
\mathfrak{D}_{\mu}\partial^{\mu}\phi^{X},
\label{Eq:EOM1}
\end{eqnarray}

\noindent
where $\mathfrak{D}_{\mu}$ denotes the covariant derivative in space-time and
in the corresponding scalar manifold. 

Along with these equations we have the Bianchi identities of the vector
fields,\footnote{By (anti-) self-duality, the Bianchi identities of the 3-form
field strengths are the equations of motion themselves.} 

\begin{equation}
dF^{i}=0. 
\end{equation}

It is also convenient to write the equations of motion of the 1- and 2-forms
in their dual form

\begin{eqnarray}
\star \mathcal{E}_{r}{}^{\mu\nu}
& = & 
\nabla_{\rho}\left(G_{rs}H^{s\ \rho \mu\nu}\right)
+\tfrac{1}{8}c_{r\
  ij}\frac{\epsilon^{\mu\nu\rho\sigma\alpha\beta}}{\sqrt{|g|}}
F^{i}{}_{\rho\sigma}F^{j}{}_{\alpha\beta},
\\
& & \nonumber \\
-\tfrac{1}{8}\star \mathcal{E}_{i}{}^{\nu}
& = & 
\nabla_{\mu}(L_{r}c^{r}{}_{ij}F^{j\ \mu\nu})
+G_{rs}c^{s}{}_{ij}H^{r\ \alpha\beta\nu}F^{j}{}_{\alpha\beta}.
\end{eqnarray}

For vanishing fermions, the supersymmetry transformations of the fermion
fields are given by\footnote{Our conventions on the spinors and gamma matrices
  are explained in Appendix \ref{app-gamma}}

\begin{eqnarray}
\delta_{\epsilon} \psi^{A}_{\mu}
& = & 
\mathcal{D}_{\mu}\epsilon^{A}
-\tfrac{1}{4}\slashed{\mathcal{H}}_{\mu}\epsilon^{A},
\\
& & \nonumber \\
\delta_{\epsilon}\chi^{M\ A}
& = & 
\tfrac{1}{2}\left[\slashed{\partial}\varphi^{\underline{\alpha}}
V_{\underline{\alpha}}{}^{M}
+\tfrac{1}{6}\slashed{\mathcal{H}}^{M}\right]\epsilon^{A},
\\
& & \nonumber \\
\delta_{\epsilon} \lambda^{i\ A}
& = &
-\tfrac{1}{2\sqrt{2}}\slashed{F}^{i}\epsilon^{A},
\\
& & \nonumber \\
\delta_{\epsilon}\Psi^{\bf{a}}
& = & 
i\slashed{\partial}\phi^{X}V^{{\bf{a}}A}{}_{X}\epsilon_{A},
\end{eqnarray}

\noindent
where $\mathcal{D}_{\mu}$ is the space-time and $\mathrm{Sp}(1)$ covariant
derivative

\begin{equation}
\mathcal{D}_{\mu}\epsilon^{A}
=
\nabla_{\mu}\epsilon^{A}+A_{\mu}{}^{A}{}_{B}\epsilon^{B}
=
\left(\partial_{\mu}
-\tfrac{1}{4}\omega_{\mu}{}^{ab}\gamma_{ab}\right)\epsilon^{A}
+A_{\mu}{}^{A}{}_{B}\epsilon^{B},
\end{equation}

\noindent
and $A_{\mu}{}^{A}{}_{B}
\equiv 
\partial_{\mu}\phi^{X}A_{X}{}^{A}{}_{B}$ is the pullback of the connection.

\section{Supersymmetric configurations}
\label{sec-configurations}

In this section we are going to identify all the supersymmetric configurations
of the theories that we have just introduced. First, in Section~\ref{Necessary
  conditions} we are going to find the necessary conditions that a field
configuration has to satisfy in order for the Killing Spinor Equations (KSEs)
to admit at least a solution (a Killing spinor). In a second stage, in
Section~\ref{sec-sufficient} we will show that these conditions are also
sufficient and we will explicitly determine the form of the Killing spinor.

\subsection{Necessary conditions}
\label{Necessary conditions}

We assume that we have certain purely bosonic field configuration which admits
a Killing spinor $\epsilon^{A}$. By definition, this means that every field is
invariant under the supersymmetry transformation generated by $\epsilon^{A}$
and, in particular, 

\begin{equation}
\delta_{\epsilon} f=0,
\end{equation}

\noindent
for every fermion $f$ of the theory and for $\epsilon^{A}$. These equations
are, by definition, the KSEs of the theory.\footnote{The supersymmetry
  transformations of the bosonic fields, being proportional to the fermionic
  fields, which vanish by assumption, are trivially satisfied.}

In order to find useful information from these fermionic equations, we will
use the ``bilinear method'' pioneered in Ref.~\cite{Gauntlett:2002nw}. Given
a Killing spinor $\epsilon^{A}$ we can construct an associated vector and a
triplet of 3-form bilinears, as explained in Appendix~\ref{app-gamma}:

\begin{equation}
l_{\mu}
\equiv
\bar{\epsilon}^{A}\gamma_{\mu}\epsilon_{A},
\quad 
W^{x}{}_{\mu\nu\rho}
\equiv
i(\sigma^{x})^{B}_{\ A}\bar{\epsilon}^{A}\gamma_{\mu\nu\rho}\epsilon_{B}.
\end{equation}

\noindent
The properties of these bilinears are described in
Appendix~\ref{Bilinears}. In particular, the triplet $W^{x}$ is anti-self dual
and $l$ is null and transverse to $W$:

\begin{equation}
\star W^{x}=-W^{x},
\quad
l_{\mu}l^{\mu}=0, 
\quad 
l^{\lambda}W^{x}{}_{\mu\nu\lambda}=0.
\end{equation}

We introduce an auxiliary null vector $n_{\mu}$ satisfying

\begin{equation}
n_{\mu}n^{\mu}=0,\quad l_{\mu}n^{\mu}=1, 
\end{equation}

\noindent
and define

\begin{equation}
\mathfrak{J}^{x}{}_{\mu\nu}\equiv n^{\lambda}W^{x}{}_{\mu\nu\lambda}.  
\end{equation}

\noindent
$\mathfrak{J}^{x}$ is transverse to $l$ and $n$, self-dual in the
four-dimensional transverse space and, most importantly, with one index
raised, $\mathfrak{J}^{x\ \mu}{}_{\nu}$, it satisfies the quaternionic algebra
Eq.~(\ref{quaternion1}).

Our next task is to extract all the possible information from the KSEs by
using these bilinears. The analysis is more or less independent for each
equation and we dedicate one section to each of them.

\subsubsection{Gravitino equation}

The gravitino KSE, $\delta_{\epsilon}\psi^{A}=0$, can be written as

\begin{equation}
\label{GravitinoKSE}
\nabla_{\mu}\epsilon^{A}
=
\tfrac{1}{4}\slashed{\mathcal{H}}_{\mu}\epsilon^{A}
-A_{\mu}{}^{A}{}_{B}\epsilon^{B}.
\end{equation}

We are going to translate this spinorial equation into equations for the
spinor bilinears.  By taking their covariant derivatives and using the above
KSE, we find that $l_{\mu}$ and $W^{x}{}_{\mu\nu\rho}$ satisfy the following
identities\footnote{These identities imply that $l_{\mu}$ and
  $W^{x}{}_{\mu\nu\rho}$ are covariantly constant with respect to the
  torsionful connection $\Gamma(e)+\mathcal{H}+A$.}  

\begin{eqnarray}
\label{leq}
\nabla_{\mu}l_{\nu}
& = & 
l^{\lambda}\mathcal{H}_{\mu\nu\lambda},
\\
& & \nonumber \\
\label{weq}
\nabla_{\mu}W^{x}{}_{\nu\rho\sigma}
& = & 
3W^{x}{}_{\lambda[\rho\sigma|}\mathcal{H}_{\mu|\nu]}{}^{\lambda}
-\epsilon^{xyz}A^{y}{}_{\mu}W^{z}{}_{\nu\rho\sigma}.
\end{eqnarray}

Eq.~(\ref{leq}) implies, in particular, that $\nabla_{(\mu}l_{\nu)}=0$, so
$l_{\mu}$ is a null Killing vector. Then, let us characterize all the metrics
which allow for a null Killing vector which in general is not covariantly
constant. First, we introduce a coordinate $v$ associated to $l^{\mu}$ defined
through

\begin{equation}
l^{\mu}\partial_{\mu}=\partial_{v}.
\end{equation}

Hence, in this coordinate system, $l^{\mu}=\delta^{\mu}_{\ v}$. On the other
hand, $l_{\mu}$ is transverse to $l^{\mu}$. Let us write
$\hat{l}=l_{\mu}dx^{\mu}$. Since, in general, $d\hat{l}\neq 0$, we cannot find
a coordinate $u$ such that $\hat{l}=du$.  In addition, generically $\hat{l}$ is not
hypersurface-orthogonal, $d\hat{l}\wedge \hat{l}\neq 0$, so
$\hat{l}\neq f du$ for any function $f$ and coordinate $u$.  Therefore, we
must write in general

\begin{equation}
\hat{l}=f(du+\beta),
\,\,\,\,\
\text{where}
\,\,\,\,\,
\beta=\beta_{\underline{m}}dx^{\underline{m}},
\,\,\,\,
\underline{m}=1,2,3,4,
\end{equation}

\noindent
so $\beta$ is a 1-form on the four-dimensional space transverse to $l$ and $n$
while $f$ is just a function.  Both $f$ and $\beta$ can depend on $u$ and
$x^{\underline{m}}$ but not on $v$.  Now, since $l^{\mu}n_{\mu}=1$, the 1-form
$\hat{n}=n_{\mu}dx^{\mu}$ can be written as
 
\begin{equation}
\hat{n}=dv+Hdu+\omega,
\end{equation}

\noindent
where $H$ is a function which again can depend on $u$ and $x^{\underline{m}}$,
and $\omega$ is a 1-form in the four-dimensional space (which can also depend
on $u$).  Finally, since $n$ and $l$ are null, the metric must be given by

\begin{equation}
ds^{2}=2\hat{l}\otimes \hat{n}-f^{-1}\delta_{mn}v^{m}\otimes v^{n},
\end{equation}

\noindent
where $v^{m}$ is the Vielbein of the four-dimensional Euclidean space which we
will call, as it is customary, ``base space''. In coordinate form, the metric
reads

\begin{equation}
ds^{2}
=
2f(du+\beta)(dv+Hdu+\omega)
-f^{-1}\gamma_{\underline{m}\underline{n}}dx^{\underline{m}} dx^{\underline{n}},
\end{equation}

\noindent
where $\gamma_{\underline{m} \underline{n}}$ is the metric of the base
space.\footnote{It is related to the Vielbein by the usual expression
  \begin{equation}
\gamma_{\underline{m} \underline{n}}
= 
\delta_{mn}v^{m}{}_{\underline{m}}v^{n}{}_{\underline{n}}.
\end{equation}
} No quantity in this metric depends on the isometric null coordinate $v$.  In
order to make any further progress in our analysis, we introduce a null
Vielbein $e^{a}{}_{\mu}$:

\begin{equation}
e^{+}
=
f(du+\beta),
\quad 
e^{-}
=
dv+Hdu+\omega, 
\quad 
e^{m}=f^{-1/2}v^{m},
\end{equation}

\noindent
and the inverse Vielbein is

\begin{equation}
e_{+}
=
f^{-1}(\partial_{u}-H\partial_{v}), 
\quad 
e_{-}
=
\partial_{v},
\quad 
e_{m}
=
f^{1/2}v_{m}-f^{1/2}\beta_{m}\partial_{u}
-f^{1/2}(\omega_{m}-\beta_{m})\partial_{v},
\end{equation}

\noindent
where $\beta_{m}\equiv v_{m}{}^{\underline{m}}\beta_{\underline{m}}$, and the
same for $\omega_{m}$. Note that $e^{+}=\hat{l}$, $e_{-}=\hat{n}$ and
$e_{+}{}^{\mu}=n^{\mu}$, $e_{-}{}^{\mu}=l^{\mu}$. The spin connection of this
Vielbein is computed in Appendix~\ref{app-curvature}.



From the condition Eq.~(\ref{leq}) we have been able to find the generic form
of the metric of a supersymmetric configuration. The next step is to exploit
the rest of information contained there in (\ref{weq}) in order to find the
general form of the field strength $\mathcal{H}$ in a supersymmetric
configuration. 

First, let us inspect the independent components of $\mathcal{H}_{abc}$ (in
the null Vielbein basis). It has four kind of components:

\begin{equation}
\mathcal{H}_{+-m}, 
\quad 
\mathcal{H}_{+mn}, 
\quad 
\mathcal{H}_{-mn}, 
\quad 
\mathcal{H}_{mnp}.
\end{equation}

\noindent
On account on the anti-self-duality of  $\mathcal{H}$:

\begin{equation}
\mathcal{H}_{mnp}=\tilde\epsilon^{mnpq}\mathcal{H}_{+-q},
\label{dualmnr}
\end{equation}

\noindent
where the Levi-Civita symbol of the transverse space is given by 

\begin{equation}
\label{LeviCivita4}
\tilde{\epsilon}^{mnrq}=\epsilon^{mnrq+-}.
\end{equation}

\noindent
Furthermore, considered as 2-forms in this four-dimensional space,
$\mathcal{H}_{+mn}$ and $\mathcal{H}_{-mn}$ are, respectively, self-dual and
anti-self-dual:
 
\begin{equation}
\tilde{\star} \mathcal{H}_{+mn}
=
+\mathcal{H}_{+mn}, 
\quad 
\tilde{\star} \mathcal{H}_{-mn}=-\mathcal{H}_{-mn}.
\end{equation}

All these conditions reduce the list of independent components of
$\mathcal{H}$.  Let us proceed with the computation of the independent
components.  In flat indices, Eq.~(\ref{leq}) can be written as

\begin{equation}
de^{+}=\mathcal{H}_{ab-}e^{a}\wedge e^{b},
\end{equation}

\noindent
and we get the following relation between $\mathcal{H}$ and the spin
connection:

\begin{equation}
\label{H-ident}
\omega_{ab-}+\mathcal{H}_{ab-}=0.
\end{equation}

\noindent
Automatically, this relation gives us the components $\mathcal{H}_{-mn}$ and
$\mathcal{H}_{+-m}$, and, by using the duality relation Eq.~(\ref{dualmnr}),
we also obtain $\mathcal{H}_{mnp}$.  Hence, it only remains to find
$\mathcal{H}_{+mn}$, or, equivalently, its self-dual part.  In order to do so,
let us note that Eq.~(\ref{weq}) can be written as

\begin{equation}
\mathcal{D}_{a} W^{x}{}_{bcd}=3W^{x}{}_{e[cd|}\mathcal{H}_{a|b]}{}^{e},
\label{weq2}
\end{equation}

\noindent
where now $\mathcal{D}_{a}$ is also the $\mathrm{Sp}(1)$ covariant
derivative.  Then, we take into account that, since $W^{x}$ is anti-self-dual
and transverse to $e^{+}$, the only non-vanishing components are
$W^{x}{}_{+mn}=\mathfrak{J}^{x}{}_{mn}$.  By using also Eq.~(\ref{H-ident}) we
see that Eq.~(\ref{weq2}) is equivalent to

\begin{equation}
\mathcal{D}_{a}\mathfrak{J}^{x}{}_{mn}
=
2\mathfrak{J}^{x}{}_{[m|p}\mathcal{H}_{a|n]}{}^{p}.
\label{Jrelation}
\end{equation}

By using again (\ref{H-ident}), the component $a=-$, gives us 
$\partial_{v}\mathfrak{J}^{x}{}_{mn}=0$.\footnote{We are advancing that $A_{-}^{x}=0$.}
 On the other hand, the component $a=+$ gives us the following relation:
 
\begin{equation}
\mathcal{H}^{+}_{+mn}+\omega^{+}_{+mn}
=
\mathfrak{J}^{z}{}_{mn}
\left(\tfrac{1}{16}f^{-1}\epsilon^{xyz}
\partial_{u}\mathfrak{J}^{x}{}_{rs}\mathfrak{J}^{y\ rs}
-\tfrac{1}{2}A_{+}^{z}\right).
\end{equation}

Since $\mathcal{H}_{+mn}^{+}=\mathcal{H}_{+mn}$, we have determined the
general form of all the components of $\mathcal{H}$.

From the gravitino KSE we can also obtain information about the base space.
In order to simplify the notation, let us introduce the following derivative
operator acting on the p-form $\alpha$ \cite{Gutowski:2003rg}

\begin{equation}
D\alpha \equiv \tilde{d} \alpha-\beta\wedge \dot{\alpha},
\end{equation}

\noindent
where $\tilde{d}$ is the exterior derivative in the base space and where
$\dot{\alpha}$ denotes the derivative with respect to $u$ in the coordinate
basis. For example, for a 1-form:

\begin{equation}
\alpha=\alpha_{\mu}dx^{\mu}\Rightarrow \dot{\alpha}
\equiv 
\partial_{u}\alpha_{\mu} dx^{\mu}.
\end{equation}

Note that the components of $\dot{\alpha}$ in the Vielbein basis are given by
$\dot{\alpha}_{a}=e_{a}{}^{\mu}\partial_{u}\alpha_{\mu}$ (in the case of a
1-form).  The operator $D$ satisfies the identity

\begin{equation}
D^{2}\alpha=-D\beta\wedge \dot{\alpha}.
\end{equation}

By using this operator, the full exterior derivative is given by

\begin{equation}
d\alpha = D\alpha+f^{-1}e^{+}\wedge \dot{\alpha}.
\end{equation} 

Coming back to the issue of interest, we know that the structures
$\mathfrak{J}^{x\ m}{}_{n}$ satisfy the quaternionic algebra. However, these
are not the natural quaternionic structures of the base space, since they must
be defined with respect to the Vielbein $v_{m}{}^{\underline{m}}$.  Therefore,
we define the complex structures in the coordinate basis as

\begin{equation}
\mathsf{J}^{x\ \underline{m}}{}_{\underline{n}}
\equiv 
v_{m}{}^{\underline{m}}\mathfrak{J}^{x\  m}{}_{n}v^{n}{}_{\underline{n}}.
\end{equation}

\noindent
Now the indices are raised and lowered with
$\gamma_{\underline{m}\underline{n}}$ instead of $g_{\mu\nu}$:

\begin{equation}
\mathsf{J}^{x}{}{}_{\underline{m}\underline{n}}
=\gamma_{\underline{m}\underline{p}}
\mathsf{J}^{x\ \underline{p}}{}_{\underline{n}}, 
\quad 
\mathsf{J}^{x\  \underline{m}\underline{n}}
=
\mathsf{J}^{x\ \underline{m}}{}_{\underline{p}}
\gamma^{\underline{p}\underline{n}}.
\end{equation}

These relations imply, in particular, that, as 2-forms, the
$\mathfrak{J}^{x}$s and $\mathsf{J}^{x}$s are related by 

\begin{equation}
\mathsf{J}^{x}=-f\mathfrak{J}^{x}, 
\end{equation}

\noindent
while in the corresponding Vielbein basis their components are related by

\begin{equation}
\mathsf{J}^{x}{}{}_{mn}\big|_{v}=-\mathfrak{J}^{x}{}{}_{mn}\big|_{e}.  
\end{equation}

Now, we can express $W^{x}$ in terms of $\mathsf{J}^{x}$ as

\begin{equation}
W^{x}=e^{+}\wedge \mathfrak{J}^{x}=-f^{-1}e^{+}\wedge \mathsf{J}^{x}.
\end{equation}

From Eq.~(\ref{weq}), it follows that 

\begin{equation}
dW^{x}+\epsilon^{xyz}A^{y}\wedge W^{z}=0. 
\end{equation}

\noindent
Then, this equation implies that

\begin{equation}
\label{dJ}
\tilde{d}\mathsf{J}^{s}
+\epsilon^{xyz}\mathsf{A}^{y}\wedge \mathsf{J}^{z}
=
\partial_{u}\left(\beta\wedge \mathsf{J}^{x}\right)
+\epsilon^{xyz}A^{y}{}_{u} \beta\wedge\mathsf{J}^{z}, 
\quad (D\beta)^{+}=0,
\end{equation}

\noindent
where $\mathsf{A}^{y}=\tilde{d}\phi^{X}A^{y}{}_{X}$ is the pullback of the
$\mathrm{Sp}(1)$ connection onto the base space, and
$A^{y}{}_{u}=\partial_{u}\phi^{X}A_{X}^{y}$.  Note that in the cases $\beta=0$
or $u$-independent, the first equation tells us that $\mathsf{J}^{x}$ is
covariantly closed in the base space with respect to the $\mathrm{Sp}(1)$
connection. However, in a quaternionic-K\"ahler manifold the complex structures
$\mathsf{J}^{x}$ must be not only covariantly closed, but covariantly
constant. Indeed, if we use the equation (\ref{Jrelation}) we see that

\begin{equation}
\begin{aligned}
\tilde{\nabla}_{r}\mathsf{J}^{x}{}_{mn}
+\epsilon^{xyz}\mathsf{A}^{y}{}_{r}\mathsf{J}_{mn}^{z}
= &
\beta_{r}\epsilon^{xyz}A_{u}{}^{y}\mathsf{J}_{mn}^{z}
+\beta_{r}\partial_{u}
\mathsf{J}^{x}{}_{mn}-\delta_{r[m}\mathsf{J}^{x}{}_{n]s}\dot{\beta}_{s}
\\
& \\
&+\mathsf{J}^{x}{}_{r[m}\dot{\beta}_{n]}-2\mathsf{J}^{x}{}_{[m|s}U_{|n]rs},
\label{covariantJ}
\end{aligned}
\end{equation}

\noindent
where 

\begin{equation}
U_{nrs}
\equiv
-\dot{v}_{n[r}\beta_{s]}+\dot{v}_{s[r}\beta_{n]}-\dot{v}_{r[n}\beta_{s]},
\end{equation}

\noindent
so that $\mathsf{J}^{x}{}_{mn}$ is actually $\mathrm{Sp}(1)$ covariantly
constant in the cases $\beta=0$ or $u$-independent. Observe that this does not
mean that the base space is quaternionic-K\"ahler, because, precisely for
$d=4$ dimensions, the definition of a quaternionic-K\"ahler space is
different.\footnote{The holonomy is $\mathrm{Sp}(1)\times \mathrm{Sp}(1)\sim
  \mathrm{SO}(4)$ and is not special anymore. Therefore, it cannot be used to
  characterize these spaces. Instead, it is required that they are Einstein
  and a self-dual Weyl tensor. } In absence of hypermultiplets, the space is
hyperK\"ahler.

Let us summarize our results so far: in a supersymmetric configuration, the
metric and the 3-form field strength $\mathcal{H}$ are given by

\begin{eqnarray}
ds^{2}
& = &
2f(du+\beta)(dv+Hdu+\omega)
-f^{-1}\gamma_{\underline{m}\underline{n}}dx^{\underline{m}}dx^{\underline{n}},
\\
& & \nonumber \\
\mathcal{H} 
& = & 
\tfrac{1}{2}f^{-1}e^{+}\wedge e^{-}\wedge\left(Df-f
  \dot{\beta}\right)+\tfrac{1}{2}fe^{-}\wedge D\beta
\nonumber \\
& & \nonumber \\
& &
-\tfrac{1}{2}\tilde{\star}
\left(Df^{-1}+f^{-1}\dot{\beta}\right)
+e^{+}\wedge
\left[
f^{-2}
\left(-\psi+\tfrac{1}{2}\mathsf{J}^{x}A^{x}{}_{u}\right)
-\tfrac{1}{2}G^{+}
\right],
\label{Hform}
\end{eqnarray}

\noindent
where $G$ is the 2-form

\begin{equation}
G=D\omega-\tilde{d} H\wedge \beta,
\end{equation}

\noindent
and 

\begin{equation}
\psi
= 
\frac{1}{16}\epsilon^{xyz}\dot{\mathsf{J}}^{x\ \underline{r}\underline{s}}
\mathsf{J}^{y}{}{}_{\underline{r}\underline{s}}\mathsf{J}^{z},
\end{equation}

\noindent
and where all the objects that appear in these expressions are
$v$-independent.  In addition, $\beta$ satisfies the equation

\begin{equation}
\label{eq:db+}
(D\beta)^{+}=0,
\end{equation}

\noindent
and $\gamma_{\underline{m}\underline{n}}$ is the metric manifold with
self-dual complex structures $\mathsf{J}^{x}$ which satisfy the quaternionic
algebra, and whose covariant derivative is given by Eq.~(\ref{covariantJ}). 

\subsubsection{Tensorino equation}

The tensorino KSE $\delta_{\epsilon} \chi^{M\ A}=0$ reads

\begin{equation}
\left[\slashed{\partial}\varphi^{\underline{\alpha}}V_{\underline{\alpha}}{}^{M}
+\tfrac{1}{6}\slashed{\mathcal{H}}^{M}\right]\epsilon^{A}=0.
\label{tensorinoKSE}
\end{equation}

\noindent
If we contract it with $\bar{\epsilon}_{A}$, we get:

\begin{equation}
0
=
l^{\mu}\partial_{\mu}\varphi^{\underline{\alpha}}
=
\partial_{v}\varphi^{\underline{\alpha}}.
\end{equation}

\noindent
Therefore, the scalars do not depend on the isometric coordinate $v$.  Another
useful identity is obtained if we contract Eq.~(\ref{tensorinoKSE}) with
$\bar{\epsilon}_{A}\gamma^{ab}$. In this case, we obtain:

\begin{equation}
\mathcal{H}^{M}{}_{abc}l^{c}
=
l_{[a}\partial_{b]}\varphi^{\underline{\alpha}}V^{M}{}_{\underline{\alpha}}.
\end{equation}

\noindent
Decomposing the fields in this equation in their components, we find that

\begin{equation}
\mathcal{H}^{M}{}_{m+-}
=
-\tfrac{1}{2}e_{m}\varphi^{\underline{\alpha}}V^{M}{}_{\underline{\alpha}}, 
\qquad 
\mathcal{H}^{M}{}_{-mn}=0.
\end{equation}

This is all the information that we can get directly from
(\ref{tensorinoKSE}). However, if we now make use of the self-duality of
$\mathcal{H}^{M}$, we find

\begin{equation}
\mathcal{H}^{M}{}_{mnr}
=
\tfrac{1}{2}\tilde{\epsilon}^{mnrq}e_{q}
\varphi^{\underline{\alpha}}V^{M}{}_{\underline{\alpha}}.
\end{equation}

\noindent
On the other hand, we have not found any condition on
$\mathcal{H}^{M}{}_{+mn}$, but the self-duality of $\mathcal{H}^{M}$ implies
that it must be anti-self-dual in the base space:

\begin{equation}
\mathcal{H}^{M}_{+mn}\equiv I^{M}{}_{mn}=(I^{M}{}_{mn})^{-}.
\end{equation}

\noindent
Hence, we can write the self-dual 3-forms $\mathcal{H}^{M}$ as

\begin{equation}
\label{HMconfig}
\mathcal{H}^{M}
=
-\tfrac{1}{2}e^{+}\wedge e^{-}\wedge D\varphi^{\underline{\alpha}}
V^{M}{}_{\underline{\alpha}}
+\tfrac{1}{2}f^{-1}\tilde{\star} D\varphi^{\underline{\alpha}} 
V^{M}{}_{\underline{\alpha}}+e^{+}\wedge I^{M}.
\end{equation}

By taking into account the results of the previous section, we
can write the physical field strengths by using the relation

\begin{equation}
H^{r}=L^{r}\mathcal{H}+L^{r\ M}\mathcal{H}^{M},
\end{equation}

\noindent
And we find

\begin{equation}
\begin{aligned}
H^{r}
=&
\tfrac{1}{2}f^{-1}e^{+}\wedge e^{-}\wedge
\big[D(fL^{r})-\dot{\beta} f
L^{r}\big]+\tfrac{1}{2}fL^{r} e^{-}\wedge D\beta
-\tfrac{1}{2}\tilde{\star}\big[D(f^{-1}L^{r})
+f^{-1}L^{r}\dot{\beta}\big]
\\
& \\
&
+e^{+}\wedge\left\{\chi^{r}
+L^{r}\left[f^{-2}
\left(-\psi+\tfrac{1}{2}\mathsf{J}^{x}A^{x}{}_{u}\right) 
-\tfrac{1}{2}G^{+}\right]\right\},
\end{aligned}
\end{equation}

\noindent
where the quantities $L_{r}$ satisfy 

\begin{equation}
\partial_{v} L_{r}=0,
\end{equation}

\noindent
and the anti-self dual 2-forms 

\begin{equation}
\chi^{r}=L^{r\ M} I^{M},
\end{equation}

\noindent
satisfy

\begin{equation}
L_{r}\chi^{r}=0.
\end{equation}

\subsubsection{Gaugino equation}

Let us now consider the KSE of the gauginos

\begin{equation}\label{Eq:gKSE}
\delta_{\epsilon} \lambda^{i\ A}
=
-\tfrac{1}{2\sqrt{2}}\slashed{F}^{i}\epsilon^{A}=0.
\end{equation}

\noindent
By contracting this equation with $\bar{\epsilon}_{B}\gamma^{a}$, we obtain
these two equations:

\begin{eqnarray}
l^{b}F^{i}{}_{bc}
& = &
0,
\\
& & \nonumber \\
 W^{x\ abc}F^{i}{}_{bc}
& = & 
0.
\end{eqnarray}

\noindent
The first equation simply tells us that 

\begin{equation}
F^{i}{}_{a-}=0. 
\end{equation}

\noindent
Then, the second equation can be rewritten as 

\begin{equation}
\mathsf{J}^{x\  mn}F^{i}_{mn}=0, 
\end{equation}

\noindent
which means that $F^{i}_{mn}$ is anti-self-dual in the base space.  Therefore,
we can write the field strength $F^{i}$ as

\begin{equation}
F^{i}=e^{+}\wedge \theta^{i}+\tilde{F}^{i},
\end{equation}

\noindent
where $\theta^{i}$ and $\tilde{F}^{i}$ are, respectively, 1- and 2-forms in
the base space, and $\tilde{F}^{i}$ is anti-self-dual in the base space

\begin{equation}
\tilde{\star} \tilde{F}^{i}=-\tilde{F}^{i}.
\end{equation}

\subsubsection{Hyperino equation}

Finally, let us analyze the supersymmetric configurations 
for the hyperscalars. The hyperino KSE reads

\begin{equation}
\delta_{\epsilon}\Psi^{\bf{a}}
=
i\slashed{\partial}\phi^{X}V^{{\bf{a}}A}{}_{X}\epsilon_{A}=0.
\end{equation}

Contraction this equation with $\bar{\epsilon}^{B}$ just tells us that

\begin{equation}
\partial_{v}\phi^{X}=0,
\end{equation}

\noindent
as expected. On the other hand, if we contract with
$\bar{\epsilon}^{B}\gamma^{ab}$ we obtain

\begin{equation}
il_{[a}\partial_{b]}\phi^{X}V^{{\bf{a}}B}{}_{X}
+\partial_{c}\phi^{X}V^{{\bf{a}}A}{}_{X}W^{B}{}_{A\ ab}{}^{c}=0.
\end{equation}

Now, by contracting with the inverse Vielbein $V^{Y}{}_{{\bf a}B}$ and by
using Eqs.~(\ref{VVrel}) and (\ref{Jdef}), one can see that this equation is
equivalent to

\begin{equation}
\label{hyperconfig}
\partial_{m}\phi^{X}
=\mathcal{J}^{x X}{}_{Y}\partial_{n}\phi^{Y}\mathsf{J}^{x\ n}{}_{m},
\end{equation}

\noindent
which characterizes $\phi^{X}$ as a ``quaternionic map''. This is similar to
what happens in $d=5$ \cite{Bellorin:2007yp}, with the difference that in our
case the hyperscalars $\phi^{X}$ can depend on the null coordinate $u$.

\subsection{Sufficient conditions}\label{sec-sufficient}

In the previous subsection we have determined the necessary conditions for a
field configuration to be supersymmetric and we have obtained the general form
of the fields. However, this does not imply that these configurations are
actually supersymmetric and one has to make sure that there is a solution to
the KSEs. We are going to show that, when the fields take the form described
in the preceding section, there exists always a Killing spinor.

To begin with, let us consider an spinor $\epsilon^{A}$ satisfying the
following conditions

\begin{equation}
\gamma^{+}\epsilon^{A}=0\ , \quad \Pi^{xA}{}_{B}\epsilon^{B}=0,
\label{KS}
\end{equation}

\noindent
where

\begin{equation}
\Pi^{xA}{}_{B}
\equiv
\tfrac{1}{2}\left[\delta^{A}{}_{B}+\tfrac{i}{4}\slashed{\mathsf{J}}^{x}(\sigma^{x})^{A}{}_{B}\right],
\,\,\,\,\,
\text{(no summation on $x$)}.
\label{projectors}
\end{equation}

\noindent
Once the condition $\gamma^{+}\epsilon^{A}=0$ is imposed, it follows that the
$\Pi^{xA}{}_{B}$ are projectors.  Moreover, the set of all these quantities
$\left\{\gamma^{+} , \Pi^{xA}{}_{B}\right\}$ is closed under commutation, so
the conditions are consistent. We also have the relation

\begin{equation}
\Pi^{xA}{}_{B}\Pi^{yB}{}_{C}\epsilon^{C}
=
\tfrac{1}{2}\left[\Pi^{xA}{}_{C}+\Pi^{yA}{}_{C}
-|\epsilon^{xyz}|\Pi^{zA}{}_{C}\right]\epsilon^{C},
\label{sigmaJ}
\end{equation}

\noindent
so that once two of the three conditions are imposed, the third is
automatically satisfied. Since each projector reduces in $1/2$ the dimension
of the space of allowed spinors, it follows that the dimension of the space of
spinors satisfying Eqs.~(\ref{KS}) is $1/8$ of the total and there is only one
independent spinor. If these conditions guarantee that the KSEs are satisfied,
this will imply that, in general, these configurations have $1/8$ of the total
supersymmetry.  Also, note that the second condition in (\ref{KS}) is
equivalent to

\begin{equation}
i (\sigma^{x})^{A}{}_{B}\epsilon^{B}
=
\tfrac{1}{4}\slashed{\mathsf{J}}^{x}\epsilon^{A},
\quad 
i (\sigma^{x})^{A}{}_{B}\epsilon_{A}
=
-\tfrac{1}{4}\slashed{\mathsf{J}}^{x}\epsilon_{B}.
\label{Jslashed}
\end{equation}

\noindent
In addition, the first condition in (\ref{KS}) fixes a chirality in the base
space, and we obtain duality relations like

\begin{equation}
\gamma^{mn}\epsilon^{A}
=
+\tfrac{1}{2}\tilde{\epsilon}^{mnrs}\gamma_{rs}\epsilon^{A},
\end{equation} 

\noindent
thus implying that $\gamma^{mn}\epsilon^{A}$ is self-dual.

Now we are going to prove that, indeed, there is always a Killing spinor
fulfilling these properties for the configurations that satisfy the necessary
conditions identified in the previous subsection. 

We can start with the hyperino equation,
$\delta_{\epsilon}\Psi^{{\bf{a}}}=0$. First, since $\gamma^{+}\epsilon^{A}=0$
and $\partial_{-}\phi^{X}=0$, this equation only involves derivatives
$\partial_{m}\phi^{X}$. Contracting this equation with $V^{X}{}_{{\bf a}B}$,
we obtain

\begin{eqnarray}
V^{Y}{}_{{\bf{a}}B}\delta_{\epsilon}\Psi^{{\bf{a}}}
& = &
\tfrac{i}{2}\gamma^{m}\left[\partial_{m}\phi^{Y}\delta_{\epsilon}^{A}{}_{B}
+\partial_{m}\phi^{X}\mathcal{F}^{Y}{}_{X}{}^{A}{}_{B}\right]\epsilon_{A}
\\
& & \nonumber \\
& = & 
\tfrac{i}{2}\gamma^{m}\left[\partial_{m}\phi^{Y}\delta_{\epsilon}^{A}{}_{B}
+i\partial_{m}\phi^{X}\mathcal{J}^{xY}{}_{X}(\sigma^{x})^{A}{}_{B}\right]
\epsilon_{A}.
\end{eqnarray}

\noindent
Now we make use of Eq.~(\ref{Jslashed}) and we obtain

\begin{equation}
V^{y}{}_{{\bf{a}}B}\delta_{\epsilon}\Psi^{{\bf{a}}}
=
\tfrac{i}{2}\gamma^{m}
\left[\partial_{m}\phi^{Y}
-\partial_{n}\phi^{X}\mathcal{J}^{xY}{}_{X}\mathsf{J}^{x\ n}{}_{m}\right]\epsilon_{B}.
\end{equation}

\noindent
Hence, on using Eq.~(\ref{hyperconfig}) we see that the r.h.s.~vanishes, so that 
$\delta_{\epsilon}\Psi^{\bf{a}}=0$ for the spinor $\epsilon^{A}$. The gaugino equation 
is also satisfied for this spinor:

\begin{equation}
\delta_{\epsilon} \lambda^{i\ A}
=
-\tfrac{1}{2\sqrt{2}}\slashed{F}^{i}\epsilon^{A}
=
-\tfrac{1}{2\sqrt{2}}
\left(2F^{i}_{m+}\gamma^{m+} +F^{i}{}_{mn}\gamma^{mn}\right)\epsilon^{A}=0,
\end{equation}

\noindent
where we have used that $F^{i}{}_{a-}=0$, that $\gamma^{+}\epsilon^{A}=0$ and
that $F^{i}{}_{mn}$ and $\gamma^{mn}\epsilon^{A}$ have opposed chirality, so
that their contraction is zero. On the other hand, for the tensorino equation
we see that

\begin{equation}
\begin{aligned}
\delta_{\epsilon}\chi^{M\ A}
& =
\tfrac{1}{2}
\left[\slashed{\partial}\varphi^{\underline{\alpha}}V_{\underline{\alpha}}{}^{M}+\tfrac{1}{6}\slashed{\mathcal{H}}^{M}\right]\epsilon^{A}
\\
& \\
& = 
\tfrac{1}{2}\left[\gamma^{m}\partial_{m}\varphi^{\underline{\alpha}}
V_{\underline{\alpha}}{}^{M}+\mathcal{H}^{M}{}_{+-m}\gamma^{+-m}
+\tfrac{1}{6}\mathcal{H}^{M}{}_{mnr}\gamma^{mnr}\right]\epsilon^{A}
\\
& \\
&
=\tfrac{1}{2}\left[\gamma^{m}
\partial_{m}\varphi^{\underline{\alpha}}V_{\underline{\alpha}}{}^{M}
+2\mathcal{H}^{M}{}_{+-m}\gamma^{+-m}\right]\epsilon^{A}
\\
& \\
& = 
\tfrac{1}{2}\left[\gamma^{m}\partial_{m}\varphi^{\underline{\alpha}}
V_{\underline{\alpha}}{}^{M}+2\mathcal{H}^{M}{}_{+-m}\gamma^{m}\right]\epsilon^{A}
\\
& \\
& = 
0.
\end{aligned}
\end{equation}

\noindent
In the second equality we have used
$\partial_{-}\varphi^{\underline{\alpha}}=0$, $\mathcal{H}^{M}_{-mn}=0$ and
the projection $\gamma^{+}\epsilon^{A}=0$, in the third equality we have used
the fact that $\mathcal{H}^{M}$ is self-dual and in the last equality we have
used Eq.~(\ref{HMconfig}).

Hence, for every spinor satisfying the projections Eqs.~(\ref{KS}), and for
the configurations obtained in the previous section, the KSEs
$\delta_{\epsilon}\Psi^{\bf{a}}=\delta_{\epsilon} \lambda^{i\
  A}=\delta_{\epsilon}\chi^{M\ A}=0$ always hold. 

Finally, we have to prove that the gravitino KSE is also satisfied by a spinor
constrained by Eqs.~(\ref{KS}). We can write the equation
$\delta_{\epsilon}\psi^{A}_{\mu}=0$ as

\begin{equation}
\partial_{a}\epsilon^{A}
=
\tfrac{1}{4}\left(\omega_{abc}+\mathcal{H}_{abc}\right)\gamma^{bc}\epsilon^{A}
-A_{a}{}^{A}{}_{B}\epsilon^{B}.
\end{equation}

\noindent
By using Eq.~(\ref{H-ident}) and $\gamma^{+}\epsilon^{A}=0$, we can simplify
the r.h.s.~of this equation to

\begin{equation}
\partial_{a}\epsilon^{A}
=\tfrac{1}{4}\left(\omega_{amn}+\mathcal{H}_{amn}\right)\gamma^{mn}\epsilon^{A}
-A_{a}{}^{A}{}_{B}\epsilon^{B}.
\end{equation}

\noindent
Taking now into account Eq.~(\ref{Jrelation}), this expression can be
rewritten as

\begin{equation}
\partial_{a}\epsilon^{A}
=
\tfrac{1}{64}\left[-\epsilon^{xyz}\partial_{a}\mathsf{J}^{x}{}_{mn}\mathsf{J}^{y\ mn}
+8A^{z}{}_{a}\right]\slashed{\mathsf{J}}^{z}\epsilon^{A}-A_{a}{}^{A}{}_{B}\epsilon^{B}.
\end{equation}

\noindent
Then, using Eq.~(\ref{Jslashed}), we get

\begin{equation}
\partial_{a}\epsilon^{A}
=-\tfrac{1}{64}\epsilon^{xyz}\partial_{a}\mathsf{J}^{x}{}_{mn}\mathsf{J}^{y\ mn}
\slashed{\mathsf{J}}^{z}\epsilon^{A}.
\end{equation}

The $v$-independence of $\mathsf{J}^{x}$ implies that of the Killing spinor.
On the other hand, we can always find a basis of the tangent space such that 
the quaternionic structures take the form

\begin{equation}
\mathsf{J}^1=v^1\wedge v^{2}-v^3\wedge v^4, \quad \mathsf{J}^{2}=v^1\wedge v^3+v^{2}\wedge v^4, \quad \mathsf{J}^3=v^1\wedge v^4-v^{2}\wedge v^3.
\end{equation}

\noindent
In particular, the components in this basis are constant \cite{Gutowski:2003rg} $\partial_{a}\mathsf{J}^{x}{}_{mn}=0$.
Therefore, in this basis any constant spinor satisfying the constraints Eqs~(\ref{KS}) also
solves the gravitino equation.

In conclusion, we have proven that all the configurations found in the
subsection~\ref{Necessary conditions} are indeed supersymmetric and they admit
a constant Killing spinor satisfying Eqs~(\ref{KS}).

\section{Supersymmetric solutions}
\label{sec-solutions}

In the previous section we have characterized the supersymmetric configurations 
of $d=6$ ungauged supergravity in terms of a number of elementary building
blocks (functions, forms, metric) satisfying certain first-order
equations. Our goal, now, is to find under which conditions they are
solutions of the equations of motion of the theory as well. The naive answer
would be to say that those conditions are, precisely, the equations of motion;
all of them. However, once we assume that the field configuration is
supersymmetric, many of the equations of motion are equivalent or
automatically solved and only a reduced number of them remain independent and
nontrivial. This is precisely the magic that one is seeking for and the reason
why finding supersymmetric solutions is, indeed, simpler.

In order to identify these independent equations of motion one can use the
integrability equations of the KSEs, which are typically proportional to
combinations of the equations of motion, or the so called {\it Killing spinor
  identities} (KSIs) \cite{Kallosh:1993wx,Bellorin:2005hy}. These can be
understood as projections over the supersymmetric configurations of the gauge
identities associated to the local supersymmetry invariance of the
theory. They are typically derived from the supergravity action assuming
invariance under local supersymmetry transformations. In this case, doing this
is not possible and we have just worked out the integrability conditions of
the KSEs.\footnote{We suspect that the KSIs may be derived from the
  pseudo-action, assuming that it can be supersymmetrized up to terms
  proportional to the self-duality constraints or by some other
  trick. However, it is not clear how to prove that the KSIs obtained in this
  way are indeed correct, except by direct comparison with those obtained from
  the integrability conditions, since nobody is actually going to find the
  required supersymmetrization of the pseudoaction.}

\subsection{Killing spinor identities}

Let us start with the gravitino KSE, which we can write as

\begin{equation}
\mathcal{D}_{\mu}\epsilon^{A}=\tfrac{1}{4}\slashed{\mathcal{H}}_{\mu}\epsilon^{A}.
\end{equation}

\noindent
Its integrability condition, which must hold for the supersymmetric
configurations that we have determined in Section~\ref{Necessary conditions}, are

\begin{equation}
\left[\mathcal{D}_{\mu},\mathcal{D}_{\nu}\right]\epsilon^{A}
=
\tfrac{1}{4}\mathcal{D}_{\mu}(\slashed{\mathcal{H}}_{\nu}\epsilon^{A})
-\tfrac{1}{4}\mathcal{D}_{\nu}(\slashed{\mathcal{H}}_{\mu}\epsilon^{A}).
\end{equation}

\noindent
The commutator in the l.h.s.~of this equation takes the value

\begin{equation}
\left[\mathcal{D}_{\mu},\mathcal{D}_{\nu}\right]\epsilon^{A}
=
-\tfrac{1}{4}R_{\mu\nu ab}\gamma^{ab}\epsilon^{A}
+\partial_{\mu}\phi^{X}\partial_{\nu}\phi^{Y}\mathcal{F}_{XY}{}^{A}{}_{B}\epsilon^{B},
\end{equation}

\noindent
leading to the identity

\begin{equation}
\tfrac{1}{4}
\left[
R_{\mu\nu ab}\gamma^{ab}
+2\nabla_{[\mu}\slashed{\mathcal{H}}_{\nu]}
+\tfrac{1}{2}\slashed{\mathcal{H}}_{[\nu}\slashed{\mathcal{H}}_{\mu]}
\right]\epsilon^{A}
-\partial_{\mu}\phi^{X}\partial_{\nu}\phi^{Y}\mathcal{F}_{XY}{}^{A}{}_{B}\epsilon^{B}=0.
\end{equation}

\noindent
If we contract on it with $\gamma^{\nu}$, we get
\begin{equation}
\tfrac{1}{4}
\left[
-2R_{\mu a}\gamma^{a}
-4\partial_{\mu}\phi^{X}\slashed{\partial}\phi^{Y}g_{XY}
+\gamma^{\nu}\left(2\nabla_{[\mu}\slashed{\mathcal{H}}_{\nu]}
+\tfrac{1}{2}\slashed{\mathcal{H}}_{[\nu}\slashed{\mathcal{H}}_{\mu]}\right)
\right]
\epsilon^{A}=0\, 
\end{equation}

\noindent
Then, after a long computation in which we make use of the gaugino KSE
(\ref{Eq:gKSE}), we rewrite this identity as

\begin{equation}
\tfrac{1}{2}
\left[
-\mathcal{E}_{ab}\gamma^{b}
-\tfrac{1}{6}L^{r}\mathcal{E}_{r\ abcd}\gamma^{bcd}
+ L^{r}\star \mathcal{E}_{r\ ab}\gamma^{b}
+C_{a}\slashed{l}
\right]
\epsilon^{A}=0,
\label{EinsteinKSI}
\end{equation}

\noindent
where we recall that the different $\mathcal{E}$-tensors represent the
equations of motion as defined in Eqs.~(\ref{Eq:EOM})-(\ref{Eq:EOM1}), and

\begin{equation}
C_{a}\equiv \tfrac{1}{2}L_{r}c^{r}{}_{ij}
\left(2l_{a}F^{i}{}_{ +r}F^{j}{}_{+}{}^{r}
+n_{a}F^{i}{}_{mn}F^{j \ mn}
-4 F^{i}{}_{ar}F^{j}{}_{+}{}^{r}\right).
\end{equation}

From Eq.~(\ref{EinsteinKSI}) we can obtain several interesting relations among
the equations of motion: if we contract it with
$\bar{\epsilon}_{A}\gamma^{cd}$ and we assume that the 2-form equations are
already satisfied we obtain

\begin{equation}
\mathcal{E}_{r}=0\Rightarrow\mathcal{E}_{a[b}l_{c]}=0.
\end{equation}

\noindent
Then, by taking into account that $l_{c}=\delta^{+}{}_{c}$ and that
$\mathcal{E}_{ab}$ is symmetric, we find that

\begin{equation}
\mathcal{E}_{a-}=\mathcal{E}_{-a}=\mathcal{E}_{am}=\mathcal{E}_{ma}=0\, .
\end{equation}

\noindent
Hence, once the 2-form equations of motion are satisfied, so are all the
components of the Einstein equation, except for the $++$ one. 

Less interesting relations can be derived for the equations of the 2-forms,
$\mathcal{E}_{r}$. For example,

\begin{equation}
L^{r}\star\mathcal{E}_{r}^{ab}W^{x}{}_{abc}
=0,
\,\,\,\,\Rightarrow\,\,\,\, 
L^{r}\star\mathcal{E}_{r\ -m}=0,
\quad 
L^{r}\star\mathcal{E}_{r\ mn}\mathsf{J}^{x\ mn}=0,
\label{ErId}
\end{equation}

\noindent
but we shall not need them since we will compute the full equations of the 
2-forms explicitly.

Let us next consider the tensorino KSE (\ref{tensorinoKSE}). Its derivative
must also also vanish, and in particular,
$\slashed{\mathcal{D}}\delta_{\epsilon}\chi^{M \ A}=0$, where
$\slashed{\mathcal{D}}=\gamma^{a}\mathcal{D}_{a}$ is the space-time and
$\mathrm{Sp}(1)$-covariant derivative. After some computations in which we
make use of the different KSEs we get the following result:

\begin{equation}
0
=
\slashed{\mathcal{D}}\delta\chi^{M \ A}
=
\tfrac{1}{2}\left[V_{\underline{\alpha}}{}^{M}\mathcal{E}^{\underline{\alpha}}
+L^{rM}\star\mathcal{E}_{r}^{ab}\gamma_{ab}\right]\epsilon^{A}.
\end{equation}

\noindent
From this equation it is evident that once the equations for the 2-forms are 
satisfied, the equations for the scalars are also satisfied,

\begin{equation}
\mathcal{E}_{r}=0,
\,\,\,\,
\Rightarrow
\,\,\,\,
 \mathcal{E}^{\underline{\alpha}}=0.
\end{equation}

\noindent
In addition, we can obtain the identities $L^{rM}\star\mathcal{E}_{r\ -m}=0$, 
and $L^{rM}\star\mathcal{E}_{r\ mn}\mathsf{J}^{x\ mn}=0$,  implying together 
with Eq.~(\ref{ErId}) that

\begin{equation}
\star\mathcal{E}_{r\ -m}=0,\qquad \star\mathcal{E}_{r\ mn}\mathsf{J}^{x\ mn}=0\, .
\end{equation}

An explicit computation shows that $\star\mathcal{E}_{r\ mn}=0$ identically
for supersymmetric configurations. Thus, the only non-vanishing components of
the 2-form equations are $\star\mathcal{E}_{r\ +m}$
and $\star\mathcal{E}_{r\ +-}$.

From the gaugino KSE (\ref{Eq:gKSE}) we can get the following interesting
identity relating some components of the vector field equations and of the
Bianchi identities,

\begin{equation}
0
=\bar{\epsilon}_{a}\mathcal{D}_{c}
\left(
\gamma^{abc}L_{r}c^{r}{}_{ij}\delta_{\epsilon}\lambda^{i\ A}\right)
=
\tfrac{1}{4\sqrt{2}}\left[l^{[a}\star\mathcal{E}_j^{b]}
-2L_{r}c^{r}{}_{ij}\ \epsilon^{abcdef}\partial_{[c}F^{i}{}_{de]}l_{f}\right].
\end{equation}

\noindent
In obtaining this identity we have made use of several results of
Section~(\ref{Necessary conditions}). 

This equation implies that $\star\mathcal{E}_{i\ -}=0$ and that, once the
Bianchi identity $(dF^{i})_{mnr}=0$ is satisfied, then $\star\mathcal{E}_{i\
  m}=0$ is also satisfied. In addition, since $F^{i}{}_{a-}=0$ and there is no
dependence in $v$, one can see that the non-vanishing components of the
Bianchi identities are $(dF^{i})_{mnr}$ and $(dF^{i})_{+mn}$.  Hence, the only
independent equations that one needs to impose are 

\begin{equation}
(dF^{i})_{mnr}=0,
\hspace{1cm}
(dF^{i})_{+mn}=0,
\hspace{1cm}
\star\mathcal{E}_{i\ +}=0.
\end{equation}

Finally, we have to determine whether the equation for the hyperscalars

\begin{equation}
\mathcal{E}^{X}=\mathcal{D}_{\mu}\partial^{\mu}\phi^{X}=0,
\end{equation}

\noindent
is satisfied.  If we take into account that $\partial_{-}\phi^{X}=0$, we can
write it as

\begin{equation}
\mathcal{E}^{X}
=
-2\omega_{-+m}\partial^{m}\phi^{X}+\mathcal{D}_{m}\partial^{m}\phi^{X}.
\end{equation}

\noindent
On the other hand, we have the relation Eq.~(\ref{hyperconfig}). By taking the 
covariant derivative there and by using that $\mathcal{J}^{x\ X}{}_{Y}$ is 
covariantly constant we obtain

\begin{equation}
\mathcal{D}_{p}\partial_{m}\phi^{X}
=\mathcal{J}^{x\ X}{}_{Y}\mathcal{D}_{p}\partial_{n}\phi^{Y}
\mathsf{J}^{x\ n}{}_{m}
+\mathcal{J}^{x X}{}_{Y}\partial_{n}\phi^{Y}\mathcal{D}_{p}\mathsf{J}^{x\ n}{}_{m}.
\end{equation}  

\noindent
Now the covariant derivative of $\mathsf{J}^{x}$ can be read from Eq.~(\ref{Jrelation}) 
and we get

\begin{equation}
\mathcal{D}_{p}\partial_{m}\phi^{X}
=
\mathcal{J}^{x X}{}_{Y}\mathcal{D}_{p}\partial_{n}\phi^{Y}
\mathsf{J}^{x\ n}{}_{m}
+\mathcal{J}^{x X}{}_{Y}\partial_{n}\phi^{Y}
\left(\mathsf{J}^{x\ n}{}_{r}\mathcal{H}_{pm}{}^{r}
+\mathsf{J}^{x\ r}{}_{m}\mathcal{H}_{p\ r}{}^{n}\right).
\end{equation} 

\noindent
Then, if we contract $m$ and $p$ we obtain

\begin{equation}
\begin{aligned}
\mathcal{D}_{m}\partial^{m}\phi^{X} 
&=
-\delta^{pm}\mathcal{D}_{p}\partial_{m}\phi^{X}
=
-\mathcal{J}^{x X}{}_{Y}\partial_{n}\phi^{Y}\mathsf{J}^{x\
  rm}\mathcal{H}_{rm}{}{}^{n}
\\
& \\
&= 
2\mathcal{J}^{x X}{}_{Y}\partial_{n}\phi^{Y}\mathsf{J}^{x\  nm}\mathcal{H}_{+-m}
=-2\partial^{m}\phi^{X}\mathcal{H}_{+-m},
\end{aligned}
\end{equation}

\noindent
where we used that $\mathcal{D}_{[p}\partial_{n]}\phi^{Y}=0$ and we recall 
that the indices of $\mathsf{J}^{x\ r}{}_{m}$ are raised and lowered with 
$+\delta^{mn}$ instead of $-\delta^{mn}$ by definition. 

Finally, we take into 
account Eq.~(\ref{H-ident}) from where we get $\mathcal{H}_{+-m}=\omega_{+m-}
=-\omega_{-+m}$. Hence, we have proven that $\mathcal{D}_{m}\partial^{m}\phi^{X}
=2\partial^{m}\phi^{X}\omega_{-+m}$, thus implying that the equation of the 
hyperscalars is automatically satisfied.

Summarizing, we have found that the only equations that we have to solve are

\begin{equation}
  \begin{array}{rclrclrcl}
\mathcal{E}_{++} & =  & 0, &
\hspace{1cm}
\star \mathcal{E}_{r\ +-} & =  & 0, &
\hspace{1cm}
\star\mathcal{E}_{r\ +m} & = & 0, \\
& & & & & & & & \\
\star\mathcal{E}_{i\ +} & =  & 0, &
\hspace{1cm}
(dF^{i})_{mnr} & = & 0, &
\hspace{1cm}
(dF^{i})_{+mn} & = & 0. \\
\end{array}
\end{equation}

\subsection{Equations of motion}

In the previous section we learned that we only have to solve the 2-form and
vector field equations, the component $++$ of the Einstein equation and the
Bianchi identities of the vector field strengths and that some components of
these equations are already satisfied. We are now going to find the form that
the remaining equations have in terms of the building blocks of the
supersymmetric solutions.

The 2-form field equations only have two non-trivially-satisfied components:

\begin{eqnarray}
\tilde{d}\left\{ L_{r}\left[ f G^{-}+f^{-1}(\mathsf{J}^{x}A^{x}{}_{u}
-2 \psi)\right]+2f \chi_{r}\right\}
& & \nonumber \\
& & \nonumber \\
-\partial_{u}\Big\{\beta\wedge\left[ L_{r}\left( f G^{-}
+f^{-1}(\mathsf{J}^{x}A^{x}{}_{u}-2 \psi)\right)+2f \chi_{r}\right]
& & \nonumber \\
& & \nonumber \\
-\tilde{\star}[D(f^{-1}L_{r})+f^{-1}L_{r}\dot{\beta}]\Big\}
+2fc_{r\ ij}\theta^{i}\wedge\tilde{F}^{j}
& = & 
0,
\\
& & \nonumber \\
D\tilde{\star}[D(f^{-1}L_{r})+f^{-1}L_{r}\dot{\beta}]
-f D\beta\wedge(L_{r} G^{-}+2\chi_{r})+c_{r\
  ij}\tilde{F}^{i}\wedge\tilde{F}^{j}
& = & 0.
\end{eqnarray}

\noindent
The Bianchi identities for the vectors have another two non-trivial sets of
components:

\begin{eqnarray}
\dot{\tilde{F}}^{i}+f\dot{\beta}\wedge\theta^{i}- D(f\theta^{i})
& = & 0,
\\
& & \nonumber \\
D\tilde{F}^{i}+fD\beta\wedge\theta^{i}
& = & 0.
\end{eqnarray}

\noindent
The equation of the vector fields only have one non-trivial component:

\begin{equation}
  c_{r\ ij}\Big[\tilde{\star} D(f^{-1}L^{r})\wedge\theta^{j}-f^{-1}D(L^{r}\tilde{\star}\theta^{j})+(L^{r}G+2\chi^{r})\wedge\tilde{F}^{j}\Big]=0,
\end{equation}

\noindent
and, finally, the only non-trivial component of the Einstein equations
reads\footnote{The component $R_{++}$ of the Ricci tensor is computed in
  Appendix~\ref{app-curvature}.}

\begin{equation}
\begin{aligned}
-\tilde{\nabla}^{2} H+\tilde{\nabla}^{m} \dot{\omega}_{m}-\beta_{m} (\ddot\omega^{m}-\partial^{m}\dot H)
-(\dot{\omega}^{m}-\partial^{m}
H)\left(2\dot{\beta}_{m}+2\dot{v}^{n}{}_{[n}\beta_{m]}+\dot{v}_{n}^{\
    \underline{r}}v_{m\underline{r}}\beta^{n}\right)
& \\
& \\
-\tfrac{1}{4} f^{2}
G^{-2}+f^{-2}\left(\psi-\tfrac{1}{2}\mathsf{J}^{x}A^{x}{}_{u}\right)^{2}+\left(\psi^{mn}-\tfrac{1}{2}\mathsf{J}^{x\
    mn}A^{x}_ u\right)G^{+}_{mn}+5 f^{-4}\dot f^{2}
& \\
& \\
-2 f^{-3}\ddot
f+\partial_{u}\left(f^{-2}\dot{v}_{m}{}^{m}\right)+f^{-2}\dot{v}_{(mn)}\dot{v}^{(mn)}-f^{2}\chi^{r}\chi_{r}-f^{-2}\dot
L_{r}\dot{L}^{r}
& \\
& \\
+2 fL_{r}c^{r}{}_{ij}\theta^{i}_{m}\theta^{j\ m}+2
g_{XY}f^{-2}\dot{\phi}^{X}\dot{\phi}^{Y}
& = 0, \\
\end{aligned}
\end{equation}

\noindent
where, for a 2-form $G=\tfrac{1}{2}G_{\mu\nu}dx^{\mu}\wedge dx^{\nu}$ we define 
$G^{2}=G_{\mu\nu}G^{\mu\nu}$. Also, in the previous expressions, the four-dimensional 
indices are raised with $+\delta^{mn}$, for example, $\omega^{m}\beta_{m}=
+\delta^{nm}\omega_{n}\beta_{m}$ and so on.

\subsection{Solving the equations}

The preceding equations are highly coupled and non-linear, and solving them is
a considerably hard task. Nonetheless, we can sketch a possible procedure
which one would ideally use in order to solve the equations. Sine the main
source of complication comes from the $u$-dependence, by demanding
independence from this coordinate the system of equations can be recast in a
triangular form and it is possible to construct explicit solutions upon choice
of a base space.

\subsubsection{Base space}

The first thing we must do is to find a base space, for which we have to find
a solution to the system of Eqs.~(\ref{eq:db+}) and (\ref{covariantJ}). The
latter can be written in a more suggestive way as follows

\begin{equation}
\hat\nabla_{r} \mathsf{J}^{x}{}_{mn}
+\epsilon^{xyz}\hat{\mathsf{A}}^{y}{}_{r}\mathsf{J}^{z}{}_{mn}=0,
\label{covariantJ2}
\end{equation} 

\noindent
where

\begin{equation}
\hat{\mathsf{A}}^{y}{}_{r}
=
\mathsf{A}^{y}{}_{r}-\beta_{r}A^{y}{}_{u}
+\tfrac{1}{8}\epsilon^{yxz}\partial_{u}\mathsf{J}^{x}{}_{rs}\mathsf{J}^{z\ rs},
\end{equation}

\noindent
and $\hat{\nabla}$ is a torsionful connection whose components
$\hat{\omega}_{mnr}$ are determined by

\begin{equation}
\left(\tilde{d}-\beta\wedge\partial_{u}
-\tfrac{1}{2}\dot{\beta}\wedge\right)v^{m}
+\hat{\omega}_{n}{}^{m}\wedge v^{n}=0.
\end{equation}

\noindent
Note that in the $u-$independent and $\beta=0$ cases this is just the usual
spin connection.  Now, in a frame in which the components
$\mathsf{J}^{x}{}_{mn}$ are constant, the Eq.~(\ref{covariantJ2}) becomes an
algebraic relation between the $\mathrm{Sp}(1)$ connection $A^{y}$ and the
self-dual part of the connection $\hat{\omega}$:

\begin{equation}
\hat{\omega}^{+}_{rmn}
=
-\tfrac{1}{6}\left(\mathsf{A}^{x}{}_{r}
-\beta_{r}A^{x}{}_{u}\right)\mathsf{J}^{x}{}_{mn}.
\end{equation}

In the case of vanishing hyperscalars, this equation simply tells us that the
connection $\hat{\omega}$ is anti-self-dual. Moreover, if there is no
dependence on $u$, $\hat{\omega}$ coincides with the spin connection and
therefore the space is hyperK\"ahler.  

\subsubsection{Simplification of the equations}\label{section:simply}

We can in general simplify the equations of motion by introducing auxiliary
quantities and further decompositions, regardless of whether we have
determined the base space.  The following decomposition of the vector fields
is useful:

\begin{equation}
\theta^{i}
=
f^{-1}(Dz^{i}-z^{i}\dot{\beta}), 
\quad 
\tilde{F}^{i}=\tilde{d}\tilde{A}^{i}-z^{i} D\beta.
\end{equation}

\noindent
In addition, we introduce the following auxiliary quantities

\begin{eqnarray}
\Sigma_{r}
&\equiv&
f(L_{r} G^{-}+2\chi_{r})+2 c_{r\ ij}z^{i}\tilde{d}\tilde{A}^{j}
-c_{r\ ij}z^{i}z^{j}D\beta+\mathsf{J}^{x}A^{x}{}_{u}-2\psi,
\label{Sigmar}\\
& & \nonumber \\
L
&\equiv&
\tilde{d}H-\dot{\omega}-d\left(f^{-1}L^{r}c_{r\ ij}z^{i}z^{j}\right)
+\partial_{u}\left(\beta f^{-1}L^{r}c_{r\ ij}z^{i}z^{j}\right).
\end{eqnarray}

\noindent
We will treat $\Sigma_{r}$ and $L$ as independent fields which once determined
can be used to obtain $\chi_{r}$, $G^{-}$ and $\tilde{d} H$.  The equations of
motion are simplified and take the form

\begin{eqnarray}
\partial_{u}\tilde{d} \tilde{A}^{i}
& = & 
0,\\
& & \nonumber \\
(\tilde{d}\tilde{A}^{i})^{+}
& = & 
0,\\
& & \nonumber \\
\Sigma_{r}^{+}-\mathsf{J}^{x}A^{x}{}_{u}+2\psi
& = & 
0,\\
& & \nonumber \\
\tilde{d}\Sigma_{r}+\partial_{u}\left[\tilde{\star}(D(f^{-1}L_{r})
+f^{-1}L_{r}\dot{\beta})-\beta\wedge\Sigma_{r}\right]
& = & 
0,\\
& & \nonumber \\
\tilde{d}\left[\tilde{\star}(D(f^{-1}L_{r})
+f^{-1}L_{r}\dot{\beta})
-\beta\wedge\Sigma_{r}
+c_{r\  ij}\tilde{A}^{i}\wedge\tilde{d}\tilde{A}^{j}\right]
& = & 
0,\\
& & \nonumber \\
c_{r\ ij}\left[-\tilde{d}\tilde{\star}
  D(f^{-1}L^{r}z^{i})+\partial_{u}\left(\beta\wedge\star
    D(f^{-1}L^{r}z^{i})\right)+\Sigma^{r}\wedge\tilde{d}\tilde{A}^{i}\right]
& = & 
0,\\
& & \nonumber \\
\tilde{\star}\left[D\tilde{\star} L-2L\wedge\star\dot{\beta}
+\tfrac{1}{2}\Sigma_{r}\wedge\Sigma^{r}-G^{+}\wedge(\mathsf{J}^{x}A^{x}{}_{u}-2\psi)\right]
& & \nonumber \\
& & \nonumber \\
+5 f^{-4}\dot f^{2}
-2 f^{-3}\ddot
f+\partial_{u}\left(f^{-2}\dot{v}_{m}{}^{m}\right)+f^{-2}\dot{v}_{(mn)}\dot{v}^{(mn)}
& & \nonumber \\
& & \nonumber \\
-f^{-2}\dot{L}_{r}\dot{L}^{r}+2 g_{XY}f^{-2}\dot{\phi}^{X}\dot{\phi}^{Y}
& = & 
0.
\end{eqnarray}

In order to find a solution, the following procedure can be followed:

\begin{enumerate}
\item First, one determines the base space, $\beta$, and the complex
  structures $\mathsf{J}^{x}$. In the presence of hyperscalars this step gets
  coupled with the rest of equations but if we truncate the hyperscalars, it
  can be carried out independently. 

\item Second, one solves the system of equations above. The equations are
  supposed to be solved in the given order, so, ideally, one would find in
  sequence $\tilde A^{i}$, $\Sigma^{r}$, $f^{-1}L_{r}$, $z^{i}$ and
  $L$. However, as we can see, these equations are not in a triangular form,
  so finding a solution is not straightforward.  

\item Finally, one has to extract the information from the auxiliary fields
  $\Sigma_{r}$ and $L$: the 2-forms $\chi_{r}$ can be obtained by using

\begin{eqnarray}
\label{chir}
\chi_{r}
 & = &
\tfrac{1}{2}\left[f^{-1}\Sigma_{r}^{-}-L_{r} G^{-}\right],
\\
& & \nonumber \\
\label{G-}
G^{-}
& = & 
f^{-1}L^{r}\left[\Sigma_{r}^{-}-2 c_{r\ ij}z^{i}\tilde{d}\tilde{A}^{j}
+c_{r\ ij}z^{i}z^{j}D\beta\right],
\end{eqnarray}

\noindent
while $\omega$ and $H$ should be determined by solving the equations

\begin{eqnarray}
\tilde{d}H-\dot{\omega}
& = & 
L+d\left(f^{-1}L^{r}c_{r\ ij}z^{i}z^{j}\right)
-\partial_{u}\left(\beta  f^{-1}L^{r}c_{r\ ij}z^{i}z^{j}\right),
\\
& & \nonumber \\
(\tilde{d}\omega)^{-}
& = & 
G^{-}+\left[\left(L+d\left(f^{-1}L^{r}c_{r\ ij}z^{i}z^{j}\right)
-\partial_{u}\left(\beta f^{-1}L^{r}c_{r\ ij}z^{i}z^{j}\right)\right)
\wedge\beta\right]^{-},
\end{eqnarray}

\noindent
where the right hand side is supposed to be known.

\end{enumerate}

We see that the main difficult of solving the equations above, apart from
determining the base space, is that they are very entangled and they do not
have the shape of a triangular system which can be solved step-by-step. 
$u$-dependent solutions to these equations were found in \cite{Bena:2011dd} in the 
case with $n_T=1$ and no vectors and hyperscalars. An extension of those results 
to the general case considered here could be carried out elsewhere.

\subsubsection*{$u-$independent solutions}

In the $u-$independent case, the base space reduces to a hyper-K\"ahler space
when the hyperscalars are truncated.  The equations of motion get the form of
a triangular system:

\begin{eqnarray}\label{Eq:u-indep1}
(\tilde{d}\tilde{A}^{i})^{+}
& = & 
0,
\\
& & \nonumber \\
\tilde{d}\Sigma_{r}& = & \label{Eq:u-indep2}
0,
\\
& & \nonumber \\
\label{Eq:u-indep3}
\Sigma_{r}^{+}
& = & 
0,
\\
& & \nonumber \\
\label{Eq:u-indep4}
\tilde{d}\left[\tilde{\star}
  \tilde{d}(f^{-1}L_{r})-\beta\wedge\Sigma_{r}+c_{r\
    ij}\tilde{A}^{i}\wedge\tilde{d}\tilde{A}^{j}\right]
& = & 
0,
\\
& & \nonumber \\
\label{Eq:u-indep5}
c_{r\ ij}\left[-\tilde{d}\tilde{\star}
  \tilde{d}(f^{-1}L^{r}z^{i})+\Sigma^{r}\wedge\tilde{d}\tilde{A}^{i}\right]
& = & 
0,
\\
& & \nonumber \\
\label{Eq:u-indep6}
\tilde{d}\tilde{\star}\tilde{d}\left(H-f^{-1}L^{r}c_{r\ ij}z^{i}z^{j}\right) +\tfrac{1}{2}\Sigma_{r}\wedge\Sigma^{r}
& = & 
0,
\\
& & \nonumber \\
\label{Eq:u-indep7}
(\tilde{d}\omega)^{-}
-f^{-1}L^{r}\left[\Sigma_{r}
-2 c_{r\ ij}z^{i}\tilde{d}\tilde{A}^{j}+c_{r\ ij}z^{i}z^{j}\tilde{d}\beta\right]
-(\tilde{d} H\wedge\beta)^{-}
& = & 
0,
\\
& & \nonumber \\
\label{Eq:u-indep8}
\chi_{r}
-\frac{1}{2f}\left[\Sigma_{r}-L_{r}L^{s}\left(\Sigma_{s}
-2 c_{s\ ij}z^{i}\tilde{d}\tilde{A}^{j}+c_{s\ ij}z^{i}z^{j}\tilde{d}\beta\right)\right]
& = & 
0,
\end{eqnarray}

These equations can be solved step-by-step (in the given order) once the base
space is determined. In the case of a hyper-K\"ahler base space (vanishing
hyperscalars) a common technique consists in assuming the existence of one
isometry, which allows to write the metric
$\gamma_{\underline{m}\underline{n}}$ in a Gibbons-Hawking form. With this
choice it is possible to solve the preceding equations explicitly, as it was
originally done in Ref.~\cite{Gauntlett:2002nw}.  For a sake of completeness
we do that next, but before so let us note that all these solutions can be
obtained by uplifting 5-dimensional solutions. Indeed, since the coordinate
$u$ is isometric in these solutions, there always exists a space-like isometry
which is a combination of the null isometry and the $u-$isometry. Dimensional
reduction along this space-like direction will produce time-like solutions of
5-dimensional supergravity. Hence, all of the $u-$independent solutions can be
obtained by uplifting the 5-dimensional time-like solutions, which are already
known \cite{Gutowski:2005id,Bellorin:2006yr,Bellorin:2007yp,Bellorin:2008we},
using the map derived in Ref.~\cite{Cano:2016rls}.

\subsubsection{Base space with one isometry}

Further simplification of the equations can be achieved in absence of
hyperscalars (so the base space is hyper-K\"ahler) by assuming, further, that
the base space has a triholomorphic isometry.\footnote{Since, as we have
  explained, the 6-dimensional $u$-independent supersymmetric solutions are
  the uplift of the 5-dimensional timelike supersymmetric solutions, this case
  is equivalent to the timelike 5-dimensional case with one additional
  triholomorphic isometry in the base space or, depending on the choice of
  compact dimension, to the $u$-independent null 5-dimensional case. The
  interest of this exercise is that it facilitates the comparison between 6-
  and 5-dimensional supersymmetric solutions.} The metric of the base space,
then, is a Gibbons-Hawking metric of the form

\begin{equation}
\gamma_{\underline{m}\underline{n}}dx^{\underline{m}}
dx^{\underline{n}}= h^{-1}(d\varphi+\chi)^{2}+h dx^{i} dx^{i}\, , \quad i=1,2,3,
\end{equation}

\noindent
where the function $h$ and the 1-form $\chi$ satisfy\footnote{The reason for
  the negative sign is that, in the conventions we are using, the complex
  structures $\mathsf{J}^{x}{}_{mn}$ are self-dual. This implies that the spin
  connection must be anti-self-dual, and this is achieved if $\star_{3}d
  h=-d\chi$. For the sake of clarity, in this subsection the symbol $d$
  denotes the exterior derivative in the 3-dimensional Euclidean space
  $\mathbb{E}^{3}$ with metric $dx^{i} dx^{i}$.}

\begin{equation}
\label{GBcondition}
\star_{3}d h=-d\chi\, .
\end{equation}

\noindent
In order to simplify the equations, let us first note that
Eq.~(\ref{Eq:u-indep2}) implies that, locally,

\begin{equation}
\Sigma_{r}=d\sigma_{r}
\end{equation}

\noindent
for some 1-forms $\sigma_{r}$. Then, if we further decompose the fields as

\begin{eqnarray}
\label{decomposition2-1}
\beta&=&-\beta_{6} h^{-1}(d\varphi+\chi)+\breve \beta\, ,\\
& & \nonumber \\
\label{decomposition2-2}
\hat{A}^{i}&=&-\phi^{i} h^{-1}(d\varphi+\chi)+\breve A^{i}\, ,\\
& & \nonumber \\
\label{decomposition2-3}
\sigma^{r}&=&-\phi^{r} h^{-1}(d\varphi+\chi)+\breve \sigma^{r}\\
& & \nonumber \\
\label{decomposition2-4}
f^{-1}L_{r}&=&h^{-1}\left[c_{r\ ij}\, \phi^{i}\phi^{j}-\beta_{6}\phi_{r}\right]+\psi_{r}\, ,\\
& & \nonumber \\
\label{decomposition2-5}
c_{r\ ij}f^{-1}L^{r}z^{j}&=&-c_{r\ ij}\, h^{-1}\phi^{r}\phi^{j}+\xi_{i}\, ,\\
& & \nonumber \\
\label{decomposition2-6}
H&=&f^{-1}L^{r}c_{r\ ij}\, z^{i}z^{j}+\frac{1}{2}h^{-1}\phi_{r}\phi^{r}+\Lambda,
\end{eqnarray}

\noindent
we obtain the following set of equations for the scalars $\beta_{6}$,
$\phi^{i}$, $\phi^{r}$, $\psi_{r}$, $\xi_{i}$ and $\Lambda$ and the
3-dimensional 1-forms $\breve{\beta}$, $\breve{A}^{i}$ and $\breve{\sigma}^{r}$:

\begin{eqnarray}
\label{eqq1}
d\breve{\beta}+\star_{3}d\beta_{6}&=&0\, ,\\
& & \nonumber \\
\label{eqq2}
d\breve{A}^{i}+\star_{3}d\phi^{i}&=&0\, ,\\
& & \nonumber \\
\label{eqq3}
d\breve{\sigma}^{r}+\star_{3}d\phi^{r}&=&0\, ,\\
& & \nonumber \\
\label{eqq4}
d\star_{3}d\psi_{r}&=&0\, ,\\
& & \nonumber \\
d\star_{3}d\xi_{i}&=&0\, ,\\
& & \nonumber \\
d\star_{3}d\Lambda&=&0\, .
\end{eqnarray}

\noindent
The functions $\psi_{r}$, $\xi_{i}$ and $\Lambda$ are harmonic on
$\mathbb{E}^{3}$, and the integrability conditions of Eqs.~(\ref{GBcondition})
and Eqs.~(\ref{eqq1})-(\ref{eqq3}) imply that the functions $h$, $\beta_{6}$,
$\phi^{i}$ and $\phi^{r}$ are also harmonic.  Now we must determine
$\omega$. It is useful to decompose it in the following way

\begin{equation}
\omega=\omega_{6} (d\varphi+\chi)+\breve{\omega},
\end{equation}

\noindent
where $\breve{\omega}$ is a 1-form in $\mathbb{E}^{3}$. In the process of
simplifying Eq.~(\ref{Eq:u-indep7}) one finds useful the following
decomposition of $\omega_{6}$

\begin{equation}
\omega_{6}
=
\hat{\omega}_{6}-2 c_{r\ ij}\frac{\phi^{r}\phi^{i}\phi^{j}}{h^{2}}+\frac{\beta_{6}\phi_{r}\phi^{r}}{h^{2}}-\frac{H\beta_{6}}{h}\, ,
\end{equation}

\noindent
and, in terms of $\hat{\omega}_{6}$, Eq.~(\ref{Eq:u-indep7}) reads

\begin{equation}\label{Eq:omega2}
d\left(\breve{\omega}-H\breve{\beta}\right)
=
\star_{3}\left[-h^{2}d\left(\frac{\hat{\omega}_{6}}{h}\right)
-2h\psi^{r} d\left(\frac{\phi_{r}}{h}\right)
+4h\xi_{i}d\left(\frac{\phi^{i}}{h}\right)
+2h\Lambda d\left(\frac{\beta_{6}}{h}\right)\right]\, .
\end{equation}

The integrability condition of this equation provides us with the equation for
$\hat{\omega}_{6}$:

\begin{equation}
0=hd\star_{3}d\left[-\hat{\omega}_{6}-\frac{\psi^{r}\phi_{r}}{h}
+2\frac{\xi_{i}\phi^{i}}{h}+\frac{\Lambda\beta_{6}}{h}\right]\, ,
\end{equation}

\noindent
whose solution can be written as

\begin{equation}
\hat \omega_{6}
=M-\frac{\psi^{r}\phi_{r}}{h}+2\frac{\xi_{i}\phi^{i}}{h}+\frac{\Lambda\beta_{6}}{h}\, ,
\quad {\rm{where}}\quad d\star_{3} d M=0\, .
\end{equation}

Taking this result into account we may rewrite Eq.~(\ref{Eq:omega2}) as

\begin{equation}
\label{Eq:w3}
d\left(\breve \omega-\tfrac{1}{2}\breve{\beta} H\right)
=
\star_{3} \left[M dh -h dM+\phi_{r} d\psi^{r}-\psi^{r} d\phi_{r}
+2\xi_{i} d\phi^{i}-2\phi^{i} d\xi_{i}+\Lambda d\beta_{6}-\beta_{6} d\Lambda\right]\, ,
\end{equation}

\noindent
whose integrability condition is now manifestly satisfied.\footnote{As it is
  well known \cite{Denef:2000nb,Bates:2003vx}, the harmonic functions
  generically considered have singularities and the integrability condition
  will not be automatically satisfied there: additional conditions on the
  integration parameters of the harmonic functions have to be met.} 

Therefore, the complete solution is determined by specifying the set of
harmonic functions $h,\beta_{6},\phi^{i},\phi^{r}, \psi_{r}, \xi_{i},\Lambda,
M$, from which it is straightforward to obtain the 1-forms $\chi$, $\breve
\beta$, $\breve{A}^{i}$, $\breve{\sigma}^{r}$ and $\breve{\omega}$ by simple
integration of Eqs.~(\ref{GBcondition}), (\ref{eqq1})-(\ref{eqq3}) and
(\ref{Eq:w3}). We get the functions $L_{r}$ and $f$ from
Eq.~(\ref{decomposition2-4}) and the condition $L_{r}L^{r}=1$.  Explicitly, we
have for $f$

\begin{equation}
\begin{aligned}
f^{-2} 
=&
\left\{h^{-1}\left[c_{r\
      ij}\phi^{i}\phi^{j}-\beta_{6}\phi_{r}\right]+\psi_{r}\right\}\left\{h^{-1}\left[c_{s\
      ij}\phi^{i}\phi^{j}-\beta_{6}\phi_ s\right]+\psi_s\right\}\eta^{rs}
\\
\\
=&
\psi_{r}\psi^{r}
+2h^{-1}\psi^{r}\left[c_{r\  ij}\,\phi^{i}\phi^{j}
-\beta_{6}\phi_{r}\right]
+h^{-2}\beta_{6}^{2}\phi_{r}\phi^{r}
-2 h^{-2}\beta_{6}c_{r\  ij}\, \phi^{r}\phi^{i}\phi^{j}\, .
\end{aligned}
\end{equation}

\noindent
Also, from Eq.~(\ref{decomposition2-5}),  we can obtain $z^{i}$,  for which we
need to solve  a linear system of  equations once an specific  $c_{r\ ij}$ has
been chosen. We can write symbolically

\begin{equation}
z^{j}=\left[c_{r\ ij}\, f^{-1}L^{r}\right]^{-1}\left(-c_{s\ il}h^{-1}\phi^s\phi^l+\xi_{i}\right)\, ,
\end{equation}

\noindent
where $\left[c_{r\ ij}\, f^{-1}L^{r}\right]^{-1}$ denotes the inverse matrix
in $ij$ indices.  Once the scalars $z^{i}$ are determined, $H$ is given by
Eq.~(\ref{decomposition2-6}), and one can compute the anti-self-dual 2-forms
$\chi_{r}$ from Eq.~(\ref{Eq:u-indep8}). Thus, we have determined all the
building blocks of the fields, and these can be written explicitly. In
particular, let us note that the vectors $A^{i}$ are given by

\begin{equation}
A^{i}
=
-du z^{i}+(d\varphi+\chi) h^{-1}(\beta_{6}z^{i}-\phi^{i})
+\breve{A}^{i}-z^{i}\breve{\beta}\, .
\end{equation}

\section{Summary}
\label{sec-summary}

In a supersymmetric configuration, the fields are $v$-independent and have the
form

\begin{eqnarray}
ds^{2} 
& = & 
2f(du+\beta)(dv+Hdu+\omega)
-f^{-1}\gamma_{\underline{m}\underline{n}}dx^{\underline{m}}
dx^{\underline{n}},
\\
& & \nonumber \\
H^{r} 
& = & 
\tfrac{1}{2}f^{-1}e^{+}\wedge e^{-}\wedge\big(D(fL^{r})-\dot{\beta} f
L^{r}\big)+\tfrac{1}{2}fL^{r} e^{-}\wedge D\beta
\nonumber \\
& & \nonumber \\
& & 
-\tfrac{1}{2}\tilde{\star}\big(D(f^{-1}L^{r})+f^{-1}L^{r}\dot{\beta}\big)
\nonumber \\
& & \nonumber \\
& & 
+e^{+}\wedge\left\{\chi^{r}+L^{r}\left[f^{-2}
\left(-\psi+\tfrac{1}{2}\mathsf{J}^{x}A^{x}{}_{u}\right) 
-\tfrac{1}{2}G^{+}\right]\right\}\, ,\\
& & \nonumber \\
F^{i} 
& = & 
e^{+}\wedge \theta^{i}+\tilde{F}^{i},
\\
& & \nonumber \\
\partial_{m}\phi^{X} 
& = & 
\mathcal{J}^{x X}{}_{Y}\partial_{n}\phi^{Y}\mathsf{J}^{x\ n}{}_{m},
\end{eqnarray}

\noindent
where

\begin{equation}
G
=
D\omega-\tilde{d} H\wedge \beta,
\quad 
\text{and}
\quad 
\psi
= 
\tfrac{1}{16}\epsilon^{xyz}\dot{\mathsf{J}}^{x\ \underline{r}\underline{s}}
\mathsf{J}^{y}{}{}_{\underline{r}\underline{s}}\mathsf{J}^{z},
\end{equation}

\noindent
and where the derivative $D$ is defined as

\begin{equation}
D\alpha =\tilde{d} \alpha-\beta\wedge \dot{\alpha}.
\end{equation}

\noindent
In addition, the quantities that appear in these expressions (the building
blocks of the supersymmetric configurations) satisfy the following properties:

\begin{equation}
\tilde{\star} \mathsf{J}^{x}=+ \mathsf{J}^{x},
\qquad 
\tilde{\star} D\beta=-D\beta, 
\qquad 
\tilde{\star}\tilde{F}^{i}=-\tilde{F}^{i}, 
\qquad 
\tilde{\star} \chi^{r}=-\chi^{r}, 
\qquad L_{r}\chi^{r}=0,
\end{equation}

\begin{equation}
\begin{aligned}
\tilde{\nabla}_{r}\mathsf{J}^{x}{}_{mn}
+\epsilon^{xyz}\mathsf{A}^{y}{}_{r}\mathsf{J}_{mn}^{z}
=&
\beta_{r}\epsilon^{xyz}A_{u}{}^{y}\mathsf{J}_{mn}^{z}
+\beta_{r}\partial_{u} \mathsf{J}^{x}{}_{mn}
-\delta_{r[m}\mathsf{J}^{x}{}_{n]s}\dot{\beta}_{s}\\
& \\
&
+\mathsf{J}^{x}{}_{r[m}\dot{\beta}_{n]}-2\mathsf{J}^{x}{}_{[m|s}U_{|n]rs},
\end{aligned}
\end{equation}

\noindent
where 

\begin{equation}
U_{nrs}
\equiv
-\dot{v}_{n[r}\beta_{s]}+\dot{v}_{s[r}\beta_{n]}-\dot{v}_{r[n}\beta_{s]}, 
\end{equation}

\noindent
$\mathsf{A}^{y}=A^{y}{}_{X}\tilde{d}\phi^{X}$ is the pullback of the
$\mathrm{Sp}(1)$ connection onto the base space, and
$A^{y}{}_{u}=A^{y}{}_{X} \partial_{u}\phi^{X}$.  Moreover, the complex
structures $\mathsf{J}^{x}$ satisfy the quaternionic algebra.  In the cases
$\beta=0$ or $u$-independent these complex structures are
$\mathrm{Sp}(1)$-covariantly constant.

The previous configurations allow for one Killing spinor $\epsilon^{A}$ which
is constant in the basis in which the complex structures $\mathsf{J}_{mn}^{z}$ are 
constant, and which satisfies

\begin{equation}
\gamma^{+}\epsilon^{A}=0\ , \qquad \Pi^{xA}{}_{B}\epsilon^{B}=0,
\end{equation}

\noindent
where

\begin{equation}
\Pi^{xA}{}_{B}
\equiv
\tfrac{1}{2}\left[\delta^{A}{}_{B}
+\tfrac{i}{4}\slashed{\mathsf{J}}^{x}(\sigma^{x})^{A}{}_{B}\right],
\end{equation}

Finally, the field equations that must be solved for these configurations are 

\begin{eqnarray}
\tilde{d}\left\{ L_{r}\left[ f G^{-}+f^{-1}(\mathsf{J}^{x}A^{x}{}_{u}
-2 \psi)\right]+2f \chi_{r}\right\}
& & \nonumber \\
& & \nonumber \\
-\partial_{u}\Big\{\beta\wedge\left[ L_{r}\left( f G^{-}
+f^{-1}(\mathsf{J}^{x}A^{x}{}_{u}-2 \psi)\right)+2f \chi_{r}\right]
& & \nonumber \\
& & \nonumber \\
-\tilde{\star}[D(f^{-1}L_{r})+f^{-1}L_{r}\dot{\beta}]\Big\}
+2fc_{r\ ij}\theta^{i}\wedge\tilde{F}^{j}
& = & 
0,
\\
& & \nonumber \\
D\tilde{\star}[D(f^{-1}L_{r})+f^{-1}L_{r}\dot{\beta}]
-f D\beta\wedge(L_{r} G^{-}+2\chi_{r})+c_{r\
  ij}\tilde{F}^{i}\wedge\tilde{F}^{j}
& = & 
0,
\\
& & \nonumber \\
\dot{\tilde{F}}^{i}+f\dot{\beta}\wedge\theta^{i}- D(f\theta^{i})
& = & 0,
\\
& & \nonumber \\
D\tilde{F}^{i}+fD\beta\wedge\theta^{i}
& = & 
0.
\\
& & \nonumber \\
c_{r\ ij}\Big[\tilde{\star} D(f^{-1}L^{r})\wedge\theta^{j}
-f^{-1}D(L^{r}\tilde{\star}\theta^{j})+(L^{r}G+2\chi^{r})\wedge\tilde{F}^{j}\Big]
& = & 
0.
\\
& & \nonumber \\
-\tilde{\nabla}^{2} H+\tilde{\nabla}^{m} \dot{\omega}_{m}
-\beta_{m} (\ddot\omega^{m}-\partial^{m}\dot H)
& & \nonumber \\
& & \nonumber \\
-(\dot{\omega}^{m}-\partial^{m}
H)\left(2\dot{\beta}_{m}+2\dot{v}^{n}{}_{[n}\beta_{m]}+\dot{v}_{n}^{\
    \underline{r}}v_{m\underline{r}}\beta^{n}\right)
& & \nonumber \\
& & \nonumber \\
-\tfrac{1}{4} f^{2}
G^{-2}+f^{-2}\left(\psi-\tfrac{1}{2}\mathsf{J}^{x}A^{x}{}_{u}\right)^{2}+\left(\psi^{mn}-\tfrac{1}{2}\mathsf{J}^{x\
    mn}A^{x}{}_u\right)G^{+}_{mn}+5 f^{-4}\dot f^{2}
& & \nonumber \\
& & \nonumber \\
-2 f^{-3}\ddot
f+\partial_{u}\left(f^{-2}\dot{v}_{m}{}^{m}\right)+f^{-2}\dot{v}_{(mn)}\dot{v}^{(mn)}-f^{2}\chi^{r}\chi_{r}-f^{-2}\dot
L_{r}\dot{L}^{r}
& & \nonumber \\
& & \nonumber \\
+2 fL_{r}c^{r}{}_{ij}\theta^{i}_{m}\theta^{j\ m}+2
g_{XY}f^{-2}\dot{\phi}^{X}\dot{\phi}^{Y}
& = & 
0.
\end{eqnarray}

A simplification of these equations is explained in
Section~\ref{section:simply}.

\section*{Acknowledgments}

The authors would like to thank P.~Meessen for many useful conversations
during the duration of this project.  This work has been supported in part by
the MINECO/FEDER, UE grants FPA2015-66793-P and FPA2015-63667-P, and by the
Spanish Research Agency (Agencia Estatal de Investigaci\'on) through the grant
IFT Centro de Excelencia Severo Ochoa SEV-2016-0597.  The work of PAC is
funded by Fundaci\'on la Caixa through a ``la Caixa - Severo Ochoa"
international pre-doctoral grant.  TO wishes to thank M.M.~Fern\'andez for her
permanent support.

\appendix

\section{Gamma matrices, spinors and bilinears}
\label{app-gamma}

\subsection{Gamma matrices and spinors in $d=6$}

We choose the mostly $-$ signature for the Minkowski metric

\begin{equation}
\eta_{a b}=\operatorname{diag}(+1,-1,-1,-1,-1,-1).
\end{equation}

\noindent
The gamma matrices are defined through the relation

\begin{equation}
\big\{\gamma_{a},\gamma_b\big\}=2\eta_{ab}.
\end{equation}

In addition to this, we will choose that they are  antisymmetric:

\begin{equation}
\gamma_{a}^{T}=-\gamma_{a}.
\end{equation}

\noindent
As usual, it is also define

\begin{equation}
\gamma_{7}=\gamma_{0}\gamma_{1}\gamma_{2}\gamma_{3}\gamma_{4}\gamma_{5}
=
\frac{1}{6!}\epsilon^{abcdef}\gamma_{a}\gamma_{b}\gamma_{c}\gamma_{d}\gamma_{e}\gamma_{f},
\end{equation}

\noindent
which satisfies that $\gamma_{7}^{2}=1$ and
$\gamma_{7}^{T}=-\gamma_{7}$. Thus, it is Hermitian and purely imaginary. We
will use the following notation for the antisymmetrized product of gamma
matrices

\begin{equation}
\gamma_{(n)}=\gamma_{a_{1}a_{2}...a_{n}}=\gamma_{[a_{1}}\gamma_{a_{2}}...\gamma_{a_{n}]},
\end{equation}

\noindent
and the duality relation reads

\begin{equation}
\gamma^{b_{1}...b_{n}}
=
\frac{(-1)^{[n/2]}}{(6-n)!}\epsilon^{b_{1}...b_{n}a_{1}...a_{6-n}}
\gamma_{a_{1}...a_{6-n}}\gamma_{7}.
\label{dualgamma}
\end{equation}

The following useful identities are  satisfied:

\begin{equation}
\gamma_{abc}\gamma_{d}\gamma^{abc}=0,
\qquad 
\gamma_{a}\gamma_{bcd}\gamma^{a}=0,
\qquad 
\gamma_{abc}\gamma_{def}\gamma^{abc}=0.
\end{equation}

\subsubsection{Reduction to five dimensions}

We want to relate the previous antisymmetric representation of the
six-dimensional $\gamma$-matrices to the five dimensional gammas.  In $d=5$
there is a unitary representation of gamma matrices, which we call
$\tilde\gamma_{i}$, $i=0,1,2,3,4$, that satisfies

\begin{equation}
\tilde{\gamma}_{0}^{\dagger}=\tilde{\gamma}_{0}, 
\quad 
\tilde{\gamma}_{i}^{\dagger}=-\tilde{\gamma}_{i}, i\neq 0.
\end{equation}

\noindent
Moreover, $\tilde{\gamma}_{4}$ is real and the rest purely imaginary. Now,
from these matrices we can construct the six-dimensional ones by means of the
following definitions

\begin{eqnarray}
\hat{\gamma}^{i}&=&\tilde \gamma^{i}\otimes \sigma^{1}, i=0,...,4\\
\hat{\gamma}^{5}&=&1\otimes i\sigma^{2}\, ,\\
\hat{\gamma}_{7}&=&\hat{\gamma}_{0}...\hat{\gamma}_{5}=1\otimes \sigma^{3},
\end{eqnarray}

\noindent
where $\sigma^{i}$ are the Pauli matrices. The matrices $\hat{\gamma}_{a}$,
$a=0,...5$, satisfy the six-dimensional Clifford algebra; However, not all of
them are antisymmetric. In order to get an antisymmetric representation we
perform the following similarity transformation

\begin{equation}
\gamma^{a}\equiv S \hat{\gamma}^{a} S^{-1},
\end{equation}

\noindent
where $S$ and its inverse are given by

\begin{equation}
S
=
\frac{1}{\sqrt{2}}\big(\tilde{\gamma}^{0}\otimes \sigma^{1}
+\tilde{\gamma}^{4}\otimes\sigma^{3}\big), 
\quad 
S^{-1}
=
\frac{1}{\sqrt{2}}\big(\tilde{\gamma}^{0}\otimes \sigma^{1}
-\tilde{\gamma}^{4}\otimes\sigma^{3}\big).
\end{equation}

Then the $\gamma^{a}$'s are an antisymmetric representation of the
six-dimensional Clifford algebra: $\gamma_{a}^{T}=-\gamma_{a}$. The explicit
relation with the five-dimensional gammas is

\begin{eqnarray}
\gamma^{0}&=&\tilde{\gamma}^{0}\otimes\sigma^{1}\, ,\\
\gamma^{i}&=&\tilde{\gamma}^{4}\tilde{\gamma}^{i}\gamma^{0}\otimes \sigma^{3}, i=1,2,3\, ,\\
\gamma^{4}&=&-\tilde{\gamma}^{4}\otimes\sigma^{1}\, ,\\
\gamma^{5}&=&1\otimes i\sigma^{2}+\tilde{\gamma}^{0}\tilde{\gamma}^{4}\otimes 1\, ,\\
\gamma_{7}&=&\tilde{\gamma}^{4}\tilde{\gamma}^{0}\otimes \sigma^{1}.
\end{eqnarray}

Note that $\gamma_{7}^{2}=1$ and it is imaginary and antisymmetric, as it should be.

\subsubsection{Majorana-Weyl symplectic Spinors}

In $\mathcal{N}=(1,0),d=6$ supergravity we use Majorana-Weyl symplectic
spinors. These are a pair spinors, $\chi^{A}$, $A=0,1$, such that they satisfy
the Weyl condition

\begin{equation}
\gamma_{7} \chi^{A}=s \chi^{A},
\end{equation}

\noindent
where $s=\pm 1$ is the chirality, and a reality condition

\begin{equation}
\big(\chi^{A}\big)^{T}=\bar{\chi}^{A},
\label{reality}
\end{equation}

\noindent
where the Dirac conjugate, $\bar{\chi}^{A}$, is defined by

\begin{equation}
\bar{\chi}_{A}=\big(\chi^{A}\big)^{\dagger}\gamma_{0}.
\end{equation}

The $\mathrm{Sp}(1)$ indices $A,B,$ can be raised and lowered in the following way

\begin{equation}
\chi^{A}
=
\epsilon^{AB}\chi_{b},
\quad 
\chi_{A}
=
\chi^{B}\epsilon_{BA},
\quad 
\epsilon_{AB}=\epsilon^{AB}=
\begin{pmatrix}
0&1\\
-1&0
\end{pmatrix}.
\end{equation}

If $\gamma_{7}\chi^{A}=s\chi^{A}$, then its Dirac conjugate satisfies 

\begin{equation}
\bar{\chi}_{A}\gamma_{7}
=
\big(\chi^{A}\big)^{\dagger}\gamma_{0}\gamma_{7}
=
-\big(\gamma_{7}\chi^{A}\big)^{\dagger}\gamma_{0}
=
-s\bar{\chi}_{A}.
\end{equation}

Therefore, if $\chi^{A}$ and $\lambda^{A}$ have, respectively, chiralities
$s_{1}$ and $s_{2}$, we get a relation for the bilinears formed with these two
spinors:

\begin{equation}
\bar{\lambda}^{A}\gamma_{(n)}\chi^{B}
=
(-1)^{n+1}s_{1}s_{2}\bar{\lambda}^{A}\gamma_{(n)}\chi^{B}.
\end{equation}

Now, by using the identities (\ref{dualgamma}), we see that the bilinears with
$n>3$ are related to those with $n\le 3$. In particular, if both spinors have
the same chirality, the only non-vanishing bilinears are those with $n=1,3$,
and if the spinors have opposed chirality, then $n=0,2$ are the only
bilinears.

\subsubsection{Fierz identities}

The antisymmetrized product of gamma matrices provide a basis for the space of
$8\times 8$ matrices. Moreover, by using the identities (\ref{dualgamma}), we
can construct the following basis:

\begin{equation}
\big\{\mathcal{O}^{i}\big\}=\big\{1,\gamma^{a}, i\gamma^{ab}, i\gamma^{abc}, 
i\gamma^{ab}\gamma_{7}, \gamma^{a}\gamma_{7},\gamma_{7}\big\},
\end{equation}

\noindent
and the dual basis

\begin{equation}
\big\{\mathcal{O}_J\big\}=\big\{1,\gamma_{a}, i\gamma_{ab}, i\gamma_{abc}, 
i\gamma_{ab}\gamma_{7}, \gamma_{7}\gamma_{a},\gamma_{7}\big\}.
\end{equation}

\noindent
The scalar product of two matrices $M$, $N$ is defined as the trace of $MN$,
and with respect to this scalar product, the basis is orthogonal:

\begin{equation}
\operatorname{Tr}\big(\mathcal{O}_J\mathcal{O}^{i}\big)=8\delta^{I}_{J}.
\end{equation}

This allows us to expand every matrix $M$ in this basis. In particular, if
$\chi$ and $\psi$ are spinors with the same chirality, $s$, and $M$ and $N$
are matrices, we get the following identity for the product of bilinears

\begin{equation}
\left(\bar{\lambda} M\chi\right)\left(\bar\psi N\varphi\right)
=
\tfrac{1}{8}\left[\bar{\lambda} M
  \gamma^{a}(1-s\gamma_{7})N\varphi\right]\left(\bar\psi\gamma_{a}\chi\right)
-\tfrac{1}{48}\left(\bar{\lambda}~ M\gamma^{abc}N\varphi\right)
\left(\bar\psi\gamma_{abc}\chi\right).
\label{fierzidentity}
\end{equation}

\subsection{Spinor Bilinears}\label{Bilinears}

Given an spinor $\kappa^{A}$, we want to construct all the possible bilinears
with it and to the determine their properties. Here we will assume that
$\kappa^{A}$ has positive chirality $\gamma_{7}\kappa^{A}=+\kappa^{A}$.
According to our previous discussion, the only bilinears that can be
constructed are a matrix of vectors and a matrix of 3-forms. Let us define
them more precisely:

\begin{equation}
V^{A}{}_{\ B\ a}\equiv \bar{\kappa}^{A}\gamma_{a}\kappa_{b},\quad W^{A}{}_{\ B\ abc}\equiv i\bar{\kappa}^{A}\gamma_{abc}\kappa_{B}.
\end{equation}

\noindent
Now, let us take both indices down in the matrix of vectors and let us use the
reality condition Eq.~(\ref{reality}) and the antisymmetry of gamma matrices:

\begin{equation}
V_{AB\ a}
=
\bar{\kappa}_{A}\gamma_{a}\kappa_{b}
=
\left(\bar{\kappa}_{A}\gamma_{a}\kappa_{b}\right)^{T}
=
\kappa_{b}^{T}\gamma_{a}^{T}\bar{\kappa}_{A}^{T}
=
-\bar{\kappa}_{b}\gamma_{a}\kappa_{A}
=
-V_{BA\ a}.
\end{equation}

\noindent
This implies that $V_{AB\ a}=\tfrac{1}{2}\epsilon_{BA}l_{a}$, or

\begin{equation}
V^{A}{}_{\ B\ a}=\tfrac{1}{2}\delta^{A}{}_{B}l_{a},
\quad 
l_{a}=\bar{\kappa}^{A}\gamma_{a}\kappa_{A}. 
\end{equation}

Hence, we only have one vector field. On the other hand, it can be shown that,
with both indices down, $W_{AB\ abc}$ is symmetric:

\begin{equation}
W_{AB\ abc}=W_{BA\ abc}.
\end{equation}

Now, since the indices are raised with $\epsilon^{AB}$, this implies that the
matrix of 3-forms is traceless, $W^{A}{}_{A\ abc}=0$.  Moreover, we can
compute the complex conjugate of $W$:

\begin{equation}
\left(W^{A}{}_{B\ abc}\right)^{*}
=
\left(i\bar{\kappa}^{A}\gamma_{abc}\kappa_{B}\right)^{\dagger}=W^{B}{}_{A\ abc}.
\end{equation}

\noindent
This means that the matrix $W^{A}{}_{B}$ is hermitian and
traceless. Therefore, we can expand it as a linear combination of Pauli
matrices

\begin{equation}
W^{A}{}_{B\ abc}
=
\tfrac{1}{2}\left(\sigma^{x}\right)^{A}{}_{B}W^{x}{}_{abc},
\,\,\,
\Leftrightarrow 
\,\,\,
W^{x}{}_{abc}=\left(\sigma^{x}\right)^{B}{}_{A}W^{A}{}_{B\ abc}.
\end{equation}

Notice that the components $W^{x}$ are real, despite $W^{A}{}_{B}$ can be
complex. Finally, we can compute the dual of the 3-form matrix

\begin{eqnarray}
\star W^{A}{}_{B\ abc}
=
\frac{1}{3!}\varepsilon_{abc}{}{}^{def}W^{A}{}_{B\ def}
=
\frac{i}{3!}\bar{\kappa}^{A}\varepsilon_{abc}{}^{def}\gamma_{def}\kappa_{B}
=
-i\bar{\kappa}^{A}\gamma_{abc}\gamma_{7}\kappa_{B}
=
-W^{A}{}_{B\ abc}.
\end{eqnarray}

Therefore, $W$ is a matrix of anti-self-dual 3-forms. We have concluded our
analysis: from a spinor $\kappa^{A}$ we can construct one vector $l_{a}$ and
three real anti-self-dual 3-forms $W^{x}{}_{\ abc}$.

Now let us apply the Fierz identity Eq.~(\ref{fierzidentity}) to products of
$l_{a}$ and $W^{A}{}_{B\ abc}$ in order to extract relations between them. For
example, we apply this identity to $l_{a}l_{b}$ and we get

\begin{equation}
l_{a}l_{b}
=
\tfrac{1}{8}\bar{\kappa}^{A}\gamma_{a}\gamma_{c}\gamma_{b}\kappa_{A}l^{c}
+\tfrac{i}{48}\bar{\kappa}^{A}\gamma_{a}\gamma^{cde}\gamma_{b}\kappa_{b}W^{B}{}_{A\ cde}.
\end{equation}

\noindent
After the calculation of every term and simplification, we have 

\begin{equation}
l_{a}l_{b}
=
-\tfrac{1}{6}\eta_{ab}l^{2}
+\tfrac{1}{6}W^{A}{}_{B\ acd}W^{B}{}_{A}{}^{cd}{}_{b}.
\end{equation}

Then, if we contract $a$ and $b$ with $\eta^{ab}$, we have that $W^{A}{}_{B\
  acd}W^{B}{}_{A\ acd}=0$, and then, $l^{2}=-l^{2}=0$, so $l_{a}$ is a null
vector.

\begin{equation}
l_{a}l_{b}
=
\tfrac{1}{6}W^{A}{}_{B\ acd}W^{B}{}_{A}{}^{cd}{}_{b},
\quad 
l^{2}=0.
\end{equation}

We can do the same analysis with the product $l W$. This time, after a quite
long calculation, we get

\begin{equation}
l_{a}W^{A}{}_{B\ bcd}
=
W^{A}{}_{B\ a[bc}l_{d]}
-l^{e}W^{A}{}_{B\ e[bc}\eta_{d]a}
-iW^{C}{}_{B\  ae[b|}W^{A}{}_{C}{}^{e}{}_{|cd]}.
\end{equation}
If we now contract the indices $a$ and $b$ we get the following expression

\begin{equation}
l^{a}W^{A}{}_{B\ acd}
=
\tfrac{i}{3}W^{C}{}_{B\ ae[d|}W^{A}{}_{C}{}^{ae}{}_{|c]}=0.
\end{equation}

The last identity can be checked by using the anti-self-duality of
$W$. Therefore, $W$ is transverse to $l$. Finally, we have to compute the
product of two $W$'s. After simplification, the painful calculation yields the
following result:

\begin{equation}
\begin{aligned}
W^{A}{}_{B}{}^{abc}W^{C}{}_{D\ efg}
& =
\tfrac{1}{4}W^{A}{}_{D}{}^{abc}W^{C}{}_{B\ efg}
+\tfrac{3}{8}\delta^{A}{}_{D}\delta^{C}{}_{B}
\left(
6 l_{[e}\delta^{[a}{}_{f}\delta^{b}{}_{g]}l^{c]}
+l^hl^{[c}\varepsilon^{ab]}{}_{hefg}\right)
\\
& \\
&
+\tfrac{9i}{2}
\left(
\delta^{C}{}_{B}W^{A}{}_{D}{}^{[a}{}_{[ef}\delta^{b}{}_{g]}l^{c]}
-\delta^{A}{}_{D}W^{C}{}_{B}{}^{[a}{}_{[ef}\delta^{b}{}_{g]}l^{c]}\right)
\\
& \\
&
+\tfrac{9}{4}
\left(  
W^{A}{}_{D}{}^{[ab|}{}_{[e|}W^{C}{}_{B}{}^{|c]}{}_{|fg]}
-W^{A}{}_{D}{}^{[ab|h}W^{C}{}_{ B\ h[ef}\delta^{|c]}{}_{g]}  
\right).
\end{aligned}
\end{equation}

This is a very complex and rich identity and it can deliver a lot of
information. For example, if we contract two pairs of indices, this expression
simplifies to

\begin{equation}
W^{A}{}_{B}{}^{abc}W^{C}{}_{D\ ebc}
=
\left(
2\delta^{A}{}_{D}\delta^{C}{}_{B}
-\delta^{A}{}_{B}\delta^{C}{}_{D}\right)l^{a}l_{e}.
\end{equation}

\noindent
If we only contract two indices, the result is

\begin{equation}
\begin{aligned}
W^{A}{}_{B}{}^{abc}W^{C}{}_{D\ efc}
&
=
\tfrac{1}{2}\left(5\delta^{A}{}_{D}\delta^{C}{}_{B}
-4\delta^{A}{}_{B}\delta^{C}{}_{D}\right)
l_{[e}\delta^{[b}{}_{f]}l^{a]}
\\
& \\
&
-2i\left(\delta^{C}{}_{B}W^{A}{}_{D}{}^{[a}{}_{ef}l^{b]}
-\delta^{A}{}_{D}W^{C}{}_{D}{}^{[a}{}_{ef}l^{b]}\right)
\\
& \\
&
+ W^{A}{}_{D}{}^{[b|c}{}_{[e|}W^{C}{}_{B}{}^{|a]}{}_{|f]c}.
\end{aligned}
\label{WWtwoindices}
\end{equation}

In order to make further progress, we need to introduce another null vector,
$n_{a}$, such that
 
\begin{equation}
n^{2}=0, \quad l^{a}n_{a}=1.
\end{equation}

\noindent
Then, we introduce a 2-form, $\mathfrak{J}^{A}{}_{B}$, defined as

\begin{equation}
\mathfrak{J}^{A}{}_{B\ ab}\equiv n^{c}W^{A}{}_{B\ abc}.
\end{equation}

\noindent
By using the anti-self-duality of $W$, it can be shown that there is a
converse relation

\begin{equation}
W^{A}{}_{B\ abc}=3\mathfrak{J}^{A}{}_{B\ [ab}l_{c]}.
\end{equation}

\noindent
By construction, $\mathfrak{J}^{A}{}_{B}$ is transverse to $n$ and $l$:

\begin{equation}
n^{a}\mathfrak{J}^{A}{}_{B\ ab}=l^{a}\mathfrak{J}^{A}{}_{B\ ab}=0.
\end{equation}

\noindent
Therefore, this object has a four-dimensional character and it lives in the
four dimensions transverse to $n$ and $l$. In this space,
$\mathfrak{J}^{A}{}_{B}$ is self-dual

\begin{equation}
\tilde{\star}\mathfrak{J}^{A}{}_{B\ ab}
\equiv
\tfrac{1}{2}\tilde{\varepsilon}_{ab}{}^{cd}\mathfrak{J}^{A}{}_{B\ cd}
=
\mathfrak{J}^{A}{}_{B\ ab}, 
\end{equation}

\noindent
where 

\begin{equation}
\tilde{\varepsilon}^{abcd}\equiv \varepsilon^{abcdef}l_{e}n_{f},
\end{equation}

\noindent
is the Levi-Civita symbol in the the 4-dimensional space transverse to the two
null directions $l,n$.

Our next goal is to find the product $\mathfrak{J}^{A}{}_{B}\cdot
\mathfrak{J}^{C}{}_{D}$. With this aim, we contract Eq.~(\ref{WWtwoindices})
with $n_{a}n^{e}$, obtaining

\begin{equation}
\mathfrak{J}^{A}{}_{B}{}^{a}{}_{c}
\mathfrak{J}^{C}{}_{D}{}^{c}{}_{b}
=
-\tfrac{1}{4}
\left(\delta^{A}{}_{(D}\delta^{C}{}_{B)}
+3\delta^{A}{}_{[D}\delta^{C}{}_{B]}\right)\tilde{\delta}^{a}{}_{b}
+\tfrac{i}{2}\left(-\delta^{A}{}_{D}\mathfrak{J}^{C}{}_{B}{}^{a}{}_{b}
+\delta^{C}{}_{B}\mathfrak{J}^{A}{}_{D}{}^{a}{}_{b}\right),
\end{equation}

\noindent
where

\begin{equation}
\tilde\delta^{a}{}_{b}\equiv \delta^{a}{}_{b}-l^{a}n_ b-l_ bn^{a},
\end{equation}

\noindent
is the identity of the 4-dimensional space transverse to the two null
directions $l,n$\footnote{Lowering one index it would give the induced metric
  on that space.} and can be used as the projector onto this space.  

Finally, we contract everything with Pauli matrices, in order to express the
equation in terms of the real components 

\begin{equation}
\mathfrak{J}^{x\  a}{}_{b}
\equiv
\left(\sigma^{x}\right)^{B}{}_{A}\mathfrak{J}^{A}{}_{B}{}^{a}{}_{b}. 
\end{equation}
\noindent

\noindent
The result is

\begin{equation}
\mathfrak{J}^{x\ a}{}_{c}\mathfrak{J}^{y\ c}{}_{b}
=-\delta^{xy}\tilde{\delta}^{a}{}_{b}
+\varepsilon^{xyz}\mathfrak{J}^{z \ a}{}_{b},
\label{quaternion1}
\end{equation}

\noindent
or, hiding the spacetime indices,

\begin{equation}
\mathfrak{J}^{x}\cdot\mathfrak{J}^{y}
=
-\delta^{xy}+\varepsilon^{xyz}\mathfrak{J}^{z}.
\label{quaternion2}
\end{equation}

Hence, the objects $\mathfrak{J}^{x}$ satisfy the algebra of quaternions. 

\section{Connection and curvature components}
\label{app-curvature}

Let us consider the metric of the supersymmetric field configurations of
ungauged $\mathcal{N}=(1,0),d=6$ supergravity:

\begin{equation}
ds^{2}
=
2f(du+\beta)(dv+Hdu+\omega)
-f^{-1}\gamma_{\underline{m}\underline{n}}dx^{\underline{m}} dx^{\underline{n}}.
\end{equation}

\noindent
The components of the inverse metric are given by 

\begin{equation}
\begin{aligned}
g^{uu}
& = 
-\beta^{2}f, 
\qquad
g^{uv}=f^{-1}+f\beta^{\underline{m}}(\beta_{\underline{m}}H
-\omega_{\underline{m}}), 
\qquad g^{u\underline{m}}=f\beta^{m},
\\
& \\
g^{vv}
& = 
-fH^{2}\beta^{2}-2Hf^{-1}+2\beta^{\underline{m}}\omega_{\underline{m}}f H
-f\omega^{2}, 
\qquad g^{v\underline{m}}=f(\omega^{\underline{m}}-\beta^{\underline{m}} H),
\\
& \\
g^{\underline{m}\underline{n}}
&=
-f\gamma^{\underline{m}\underline{n}},
\end{aligned}
\end{equation}

\noindent
where the indices $\underline{m}$ and $\underline{n}$ in
$\omega^{\underline{m}},\beta^{\underline{n}}$ have been raised with
$\gamma^{\underline{m}\underline{n}}$. We introduce a Vielbein:

\begin{equation}
e^{+}=f(du+\beta),
\qquad 
e^{-}=dv+Hdu+\omega, 
\qquad 
e^{m}=f^{-1/2}v^{m},
\end{equation}

\noindent
where $v^{m}$ is a Vielbein of the metric
$\gamma_{\underline{m}\underline{n}}$. The metric then is

\begin{equation}
ds^{2}=e^{+}\otimes e^{-}+e^{-}\otimes e^{+}-\delta_{mn}e^{m}\otimes e^{n}.
\end{equation}

\noindent
and the inverse Vielbein is

\begin{equation}
e_{+}=f^{-1}(\partial_{u}-H\partial_{v}), 
\quad 
e_{-}=\partial_{v},
\quad 
e_{m}=f^{1/2}v_{m}-f^{1/2}\beta_{m}\partial_{u}-f^{1/2}(\omega_{m}
-\beta_{m})\partial_{v}.
\end{equation}

\noindent
The spin connection $\omega_{ab}=\omega_{cab}e^{c}$ is defined through
Cartan's first structure equation

\begin{equation}
de^{a}=\omega^{a}{}_{b}\wedge e^{b},
\end{equation}

\noindent
and, for the above metric and Vielbein basis it has the following
non-vanishing components

\begin{equation}
\begin{aligned}
\omega_{++m}
&=
f^{-1/2}\left[\dot{\omega}_{m}-v_{m}H\right],
\\
& \\
\omega_{+-m}
&=
\omega_{-+m}
=
\omega_{m+-}
=
\tfrac{1}{2}f^{-1/2}\left[\dot f\beta_{m}+f\dot{\beta}_{m}-v_{m}f\right],
\\
& \\
\omega_{+mn}
&=
f^{-1}\dot{v}_{[mn]}+\tfrac{1}{2}f(\tilde{d}\omega)_{mn}
-fv_{[m}H\beta_{n]}+f\dot{\omega}_{[m}\beta_{n]},
\\
& \\
\omega_{-mn}
&=
-\omega_{mn-}
=
\tfrac{1}{2}f^{2}(\tilde{d}\beta)_{mn}+f^{2}\dot{\beta}_{[m}\beta_{n]},
\\
& \\
\omega_{m+n}
&=
\tfrac{1}{2}f^{-2}\dot f
\delta_{mn}-f^{-1}\dot{v}_{(mn)}+f\tfrac{1}{2}(\tilde{d}\omega)_{mn}
-fv_{[m}H\beta_{n]}+f\dot{\omega}_{[m}\beta_{n]},
\\
& \\
\omega_{mnr}
&=
-f^{1/2}\tilde\omega_{mnr}+f^{-1/2}v_{[n}f\delta_{r]m}+f^{-1/2}\dot
f\beta_{[r}\delta_{n]m}
\\
& \\
& 
\,\,\, \,\,\,\, 
+f^{1/2}\left[-\dot{v}_{n[m}\beta_{r]}+\dot{v}_{r[m}\beta_{n]}
-\dot{v}_{m[n}\beta_{r]}\right].
\end{aligned}
\end{equation}

In these expressions, $\dot{v}_{mn}\equiv
\dot{v}_{m\underline{r}}v_{n}{}^{\underline{r}}$,
$\dot{\omega}_{m}=v_{m}{}^{\underline{r}}\partial_{u}\omega_{\underline{r}}$
and $(\tilde{d}\omega)_{mn}=2v_{[m}\omega_{n]}$, and $\tilde\omega_{mnr}$ is
the spin connection associated to the Vielbein $v^{n}$.

The curvature 2-form is defined by

\begin{equation}
R^{a}{}_{ b}
=
\tfrac{1}{2}R^{a}{}_{bcd}e^{c}\wedge e^{d}
=d\omega^{a}{}_{b}-\omega^{a}{}_{c}\wedge \omega^{c}{}_{b}.
\end{equation}

\noindent
In this work we only need to compute a component of the Ricci tensor, namely
$R_{++}=R^{a}{}_{+a+}=R^{m}{}_{+m+}$. We get the following result:

\begin{equation}
\begin{aligned}
R_{++}
=&
-\tilde{\nabla}^{2}H+\tilde{\nabla}^{m}\dot{\omega}_{m}
-\beta_{m}(\ddot{\omega}^{m}-\partial^{m}\dot{H})
\\
& \\
&
-(\dot{\omega}^{m}-\partial^{m}H)(2\dot{\beta}_{m}
+2\dot{v}^{n}{}_{[n}\beta_{m]}+\dot{v}^{n\underline{r}}v_{m\underline{r}}\beta_{n})
\\
& \\
&
+\tfrac{1}{4}f^{2}G^{2}+5f^{-4}\dot{f}^{2}-2f^{-3}\ddot{f}
+\partial_{u}(f^{-2}\dot{v}_{m}{}^{m})+f^{-2}\dot{v}_{(mn)}\dot{v}^{(mn)}.
\end{aligned}
\end{equation}

\noindent
where $\tilde{\nabla}$ stands for the covariant derivative with respect to the
connection $\tilde{\omega}_{mnr}$, and

\begin{equation}
G
\equiv 
\tilde{d} \omega-\tilde{d} H\wedge \beta+\dot{\omega}\wedge \beta
=D\omega-\tilde{d} H\wedge \beta.
\end{equation}


\renewcommand{\leftmark}{\MakeUppercase{Bibliography}}
\phantomsection

\end{document}